\documentclass{aa}
\usepackage{graphicx}
\usepackage{txfonts}
\newcommand{\Mn}[5]{\mbox{$#1\,^#2{\rm #3}^{{\rm #4}}_{\rm #5}$}}
\newcommand{\Te}{T_{\rm e}}
\newcommand{\mA}{{\rm m\AA}}
\newcommand{\Elow}{E_{\rm low}}

\newcommand{\loggfe}{\log gf\varepsilon}
\newcommand{\loge}{\log\varepsilon}
\newcommand{\logeMnN}[1]{\log\varepsilon_{\rm Mn, \odot}^{\rm NLTE}}
\newcommand{\logeMnL}[1]{\log\varepsilon_{\rm Mn, \odot}^{\rm LTE}}
\newcommand{\logeMn}[1]{\log\varepsilon_{\rm Mn, \odot}}
\newcommand{\logeFe}[1]{\log\varepsilon_{\rm Fe, \odot}}
\newcommand{\logemean}[1]{\log\varepsilon_{\rm mean}}
\newcommand{\kms}{km s$^{-1}$}
\newcommand{\SH}{S\!_{\rm H}}           
\newcommand{\SP}{S\!_{\rm P}}           

\begin{document}
\title{Formation of \ion{Mn}{i} lines in the solar
atmosphere\thanks{Research supported by the International Max Planck Research
School (IMPRS), Munich, Germany}}
\author{M. Bergemann \and T. Gehren}
\offprints{\\M. Bergemann, e-mail: mbergema@usm.lmu.de}
\institute{ Institute for Astronomy and Astrophysics, Ludwig-Maximilian
University, Scheinerstr. 1, 81679 Munich, Germany}
\date{Received date / Accepted date}
\abstract
{We present a detailed NLTE analysis of 39 \ion{Mn}{i} lines in the solar
spectrum. The influence of NLTE effects on the line formation and element
abundance is investigated.}
{Our goal is the derivation of solar $\loggfe$ values for manganese lines,
which will later be used in differential abundance analyses of metal-poor
stars.}
{The method of spectrum synthesis is employed, which is based on a solar model
atmosphere with initially specified element abundances. A manganese abundance of
$\logeMn\ = 5.47$ dex is used with the theoretical line-blanketed model
atmosphere. Statistical equilibrium calculations are carried out for the model
atom, which comprises 245 and 213 levels for \ion{Mn}{i} and \ion{Mn}{ii},
respectively. Photoionization cross-sections are assumed hydrogenic.}
{For line synthesis van der Waals broadening is calculated according to
Anstee $\&$ O'Mara's formalism. It is shown that hyperfine structure of the Mn
lines also has strong broadening effects, and that manganese is prone to NLTE
effects in the solar atmosphere. The nature of the NLTE effects and the validity
of the LTE approach are discussed in detail. The role of photoionization and
collisional interaction is investigated.}
{Maximum NLTE corrections of +0.1 dex with respect to LTE profiles are found. We
propose an absolute solar abundance of 5.36 $\pm\ 0.1$ dex. The main source of
errors in the abundance calculations is uncertain oscillator strengths.}
\keywords{Atomic data -- Line: profiles -- Sun: abundances}
\titlerunning{Manganese lines in the solar spectrum}
\authorrunning{M. Bergemann \& T. Gehren}
\maketitle
\section{Introduction}
Manganese belongs to the iron-group elements, which have very complex but
interesting atomic structure. Since lines in the synthetic spectrum of a
star are based on the adopted model atom and the transitions included, the
analysis becomes more and more complicated for elements with increasing number
of levels and electrons. Construction of the model atom, usually done within a
few days, becomes a matter of several months. It is necessary not only to
collect all existing data and compare them, but also to assess their
plausibility and necessity for the aims of current research.

These difficulties are, however, compensated by a beautiful pattern of lines in
the spectrum, which results from fine and hyperfine transitions in ions of such
elements. The \ion{Mn}{i} lines are split into several components; this is a
result of interaction of the electromagnetic field produced by the electrons
with the magnetic momentum of the nucleus. The large hyperfine structure (HFS)
of the lines overwhelms all other sources of line broadening. The width of a
single Doppler line at 8400 \AA\ in the solar atmosphere is $\sim$ 28 \mA\
versus $\sim$ 40 \mA\ of a HFS broadened line. Without question, it has to be
accurately taken into account, otherwise serious errors appear in the
calculation of element abundances (Abt \cite{Abt52}; Prochaska \& McWilliam
\cite{Proch00}; del Peloso et al. \cite{delPel05}; Vitas \& Vince \cite{VV03}).

It is interesting to study the manganese abundance in stars of different
populations. In metal-poor stars the manganese abundance was found to be
correlated with metallicity. A progressive deficiency of Mn relative to iron was
first reported by Helfer et al. (\cite{Helfer59}) and later confirmed in
analyses of other authors (Wheeler et al. \cite{Wheeler89}; Gratton
\cite{Gratton89}; Nissen et al. \cite{Nissen00}; Prochaska $\&$ McWilliam
\cite{Proch00}), right up to the most recent investigation of Sobeck et al.
(\cite{Sobeck05}). This element is not produced in quasi-static nuclear burning
processes; instead, it is accumulated from supernova explosions. Therefore, the
variation of the abundance pattern of manganese with metallicity may indicate
when and which stars (SN Ia, SN II), and which production processes of heavy
elements started to contribute to the chemical enrichment of the Galaxy.

Aside from global implications for studies of stellar nucleosynthesis, manganese
is an indicator of NLTE conditions in a stellar atmosphere. Mn does not
contribute to stellar opacity, as compared to iron, and it does not
significantly add to the pool of free electrons. More than 95\% of manganese is
ionized. Thus, \ion{Mn}{i} can be regarded as a trace ion in a stellar
atmosphere and studied under the NLTE assumption, which is necessary to describe
lines formed high in the photosphere (Mihalas $\&$ Athay \cite{Mihalas73}).
Departures from LTE for the resonance line at 5394 $\AA$ in the Sun have been
already found by Vitas \& Vince (\cite{VV05}), although the nature of the NLTE
effects was not investigated. Nissen et al. (\cite{Nissen00}) proposed that
Mn abundances might be subject to NLTE effects due to overionization of
\ion{Mn}{i} caused by a strong radiation field.

The goal of the current analysis is to investigate in detail departures from LTE
in $\ion{Mn}{i}$ and assess the plausibility of previous element abundance
estimates for the Sun. There is an unresolved inconsistency between the
abundance of Mn in the solar atmosphere (5.39 $\pm$ 0.03) (Booth et al.
\cite{Booth84b}) and in CI Chondrites (5.50 $\pm$ 0.03) (Lodders
\cite{Lodders03}). It is not yet clear whether to attribute this to an incorrect
allowance for HFS and NLTE effects in the calculation of the solar Mn abundance.
The fit of photospheric and meteoritic abundances of the reference element Si is
also very important in this context.

In Sect. 2 we give a short account of the method used and the full atomic
model of \ion{Mn}{i} including the choice of interaction processes. For
the purpose of identifying the important levels and transitions, this section
also presents two \emph{reduced} models. Section 3 explores the resulting
departure coefficients, line source functions, and the nature of the deviations
from LTE. It also introduces their variation with photoionization and collision
cross-sections. In Sect. 4 we dicuss the synthesis of the \ion{Mn}{i}
lines with due reference to HFS and NLTE results. The resulting solar manganese
abundance is discussed.
\section{NLTE calculations}
\subsection{Method}
In our current atomic model the following processes are taken into account:
photon absorption in line transitions, photoionization, excitation and
ionization by collisions with free electrons, and neutral hydrogen atoms.
All processes include their reverse reactions; in particular, bound-bound
radiative transitions are assumed to follow complete frequency redistribution.
It is known that under certain circumstances other interaction processes can
also play a role, e.g. autoionization transitions, charge-transfer reactions or
di-electronic recombination. No information about these processes is available
for manganese, and in particular, autoionization resonances could well
contribute to the depopulation of some \ion{Mn}{i} levels.

In this work NLTE atomic populations were computed with a revised version of the
program DETAIL (Butler $\&$ Giddings \cite{Butler85}), which solves the
radiative transfer and statistical equilibrium equations by the method of
accelerated lambda iteration. We use a theoretical line-blanketed model of the
solar photosphere MAFAGS-ODF, calculated with the MAFAGS code (Fuhrmann
et al. \cite{F97}). This model atmosphere uses opacity distribution functions
(ODF) taken from Kurucz (\cite{Kurucz92}). The resulting atmospheric
stratifications of temperature and pressure are similar to those given by 
other comparable models (see comparison in Grupp \cite{Grupp04}, Fig. 15). As is
the case with all other line-blanketed atmospheric models of this type so far,
we have not attempted to model the solar chromosphere.
\subsection{Models of the \ion{Mn}{i} atom}
\subsubsection{The reference model}
\label{refmod}
In our \emph{reference model} the manganese atom is constructed with 245 levels
for \ion{Mn}{i} and 213 levels for \ion{Mn}{ii}, respectively. The system is
closed with the ground state of \ion{Mn}{iii}. The energies for these levels are
taken from Sugar \& Corliss (\cite{Sugar85}). We include all levels with $n <
15$ for \ion{Mn}{i}, $n < 8$ for \ion{Mn}{ii}, and with energies of 0.03 and
1.11 eV below the respective ionization limits. Such a relatively complete model
should provide a very close coupling of the upper atomic levels to the next ion
ground state. The number of lines treated in NLTE is 1261 for \ion{Mn}{i} and
1548 for \ion{Mn}{ii}. 

Wavelengths and oscillator strengths are all taken from Kurucz's database
(Kurucz $\&$ Bell \cite{Kurucz95}). We note here that most of the lines do not
require data that are more precise (in fact it will turn out that the available
\emph{experimental} f-values are not better than those provided by
calculations). In a few cases of strong line blending in the terrestrial UV, 
we have tried to make sure that the calculated wavelengths are at the observed
positions. Hyperfine splitting of the lines is not included in the statistical
equilibrium calculations, since there is no reason to believe that the
\emph{relative} populations of these levels deviate from thermal. Moreover, for
the uppermost terms above 7 eV the fine structure is not maintained. These terms
are represented by a single level with a weighted mean of statistical weights
and ionization frequencies of their fine structure levels. A complete grotrian
diagram for \ion{Mn}{i} is available online. The majority of radiative 
transitions in \ion{Mn}{i} occurs between low excitation metastable and high
excitation levels. Also, transitions starting from the metastable levels produce
most of the lines, which we use in the spectrum analysis. Quite different from
\ion{Fe}{i}, the intercombination lines do not seem to form a tight coupling of
the multiplets. Except for the calculation of the ionization equilibrium, the
multiplets could therefore be calculated for themselves ignoring any multiplet
interaction. 

For bound-free radiative transitions we have to use hydrogen-like
photoionization cross-sections (Mihalas \cite{Mihalas78}) because calculated
data from the Opacity Project are not yet available. In our current analysis
this may be the most uncertain representation. We conclude this in analogy to
the \ion{Fe}{i} atom, for which we found that the calculated photoionization
cross-sections of Bautista (\cite{Baut97}) are orders of magnitude \emph{larger}
than hydrogenic approximations (Gehren et al. \cite{Gehren01a}).

Rates of transitions due to inelastic collisions with \ion{H}{i} atoms are
calculated according to Drawin's formula (\cite{Drawin68}, \cite{Drawin69}) in
the version of Steenbock \& Holweger (\cite{Steen84}). Drawin's cross-sections
are usually multiplied by a scaling factor $\SH$, which often takes  values
much smaller than unity (see the review by Asplund \cite{Asplund05}). Thus, we
decided to scale the rates of bound-bound and bound-free transitions by a factor
of 0.05 in our reference model. For allowed transitions, the cross-sections for
collisional excitation with electrons are calculated from the formula of van
Regemorter (\cite{Reg62}). Forbidden transitions due to collisions with
electrons are computed from the formula of Allen (\cite{Allen73}).
\subsubsection{Reduced models}
Our procedure of reducing the complexity of the model atom lies in a consecutive
exclusion of those atomic levels that are suspected to cause the secondary NLTE
effects, such as photon pumping or radiative recombination. These levels are
identified in Sect. 3.1. The purpose is to find a model that represents the
idealized case of the manganese atom, where deviations from LTE in level
populations are exclusively due to \emph{one primary process}. We will also be
able to establish a minimum model that gives similar results as the reference
one. As a by-product the time of computation could be significantly decreased.

As high-excitation levels provide an effective coupling to the continuum by
means of collisions, it is tempting to exclude them in order to minimize the
\emph{collisional} interaction of \ion{Mn}{i} and \ion{Mn}{ii}. A first reduced
model of \ion{Mn}{i} thus includes all 145 levels with an excitation potential
of less than 6.42 eV ($E_{\rm ion} = 7.43$ eV). The \ion{Mn}{i} levels with $n =
6 \ldots 15$ and all levels of \ion{Mn}{ii}, except the ground state, were
excluded from the calculations.

Since the increase to 1 eV of the energy gap between both ions is not enough to
ensure the dominance of radiative processes, and it does not lead to perceptible
changes in the distribution of level populations (see next section), a second
reduced model was calculated with only 65 levels ($n \leq 5$) for \ion{Mn}{i},
and an excitation potential of the uppermost level of 5.23 eV. Both
types of reduced models in practice eliminate the complete doublet system except
for the metastable states.

Whereas the above reduced model atoms are defined with respect to the excitation
energies of the contributing levels, other models may exclude single multiplet
systems, such as the doublet or the octet system. Note that statistical
equilibrium requires a valid representation of such levels in order to calculate
the ionization equilibrium. In the case of the doublet system it seems that it
can be completely ignored because it contains only a few levels and all of them
are highly excited. Also, there are no useful doublet lines in the solar
spectrum. This is different with the octet system, which is also only loosely
coupled to quartets and sextets. However, a significant fraction of the
\ion{Mn}{i} atoms is found here, in addition to some interesting solar lines.

Although we find that the number of levels (and lines) in both \ion{Mn}{i} and
\ion{Mn}{ii} can be significantly reduced without changing the formation of the
most important solar manganese lines, we have calculated all final results with
the complete reference model from Sect. \ref{refmod}. The reduced models are
therefore only used for direct investigation of critical processes.
\section{Results}
\subsection{NLTE level populations}
\begin{figure*}
\hbox{\resizebox{\columnwidth}{!}{\includegraphics{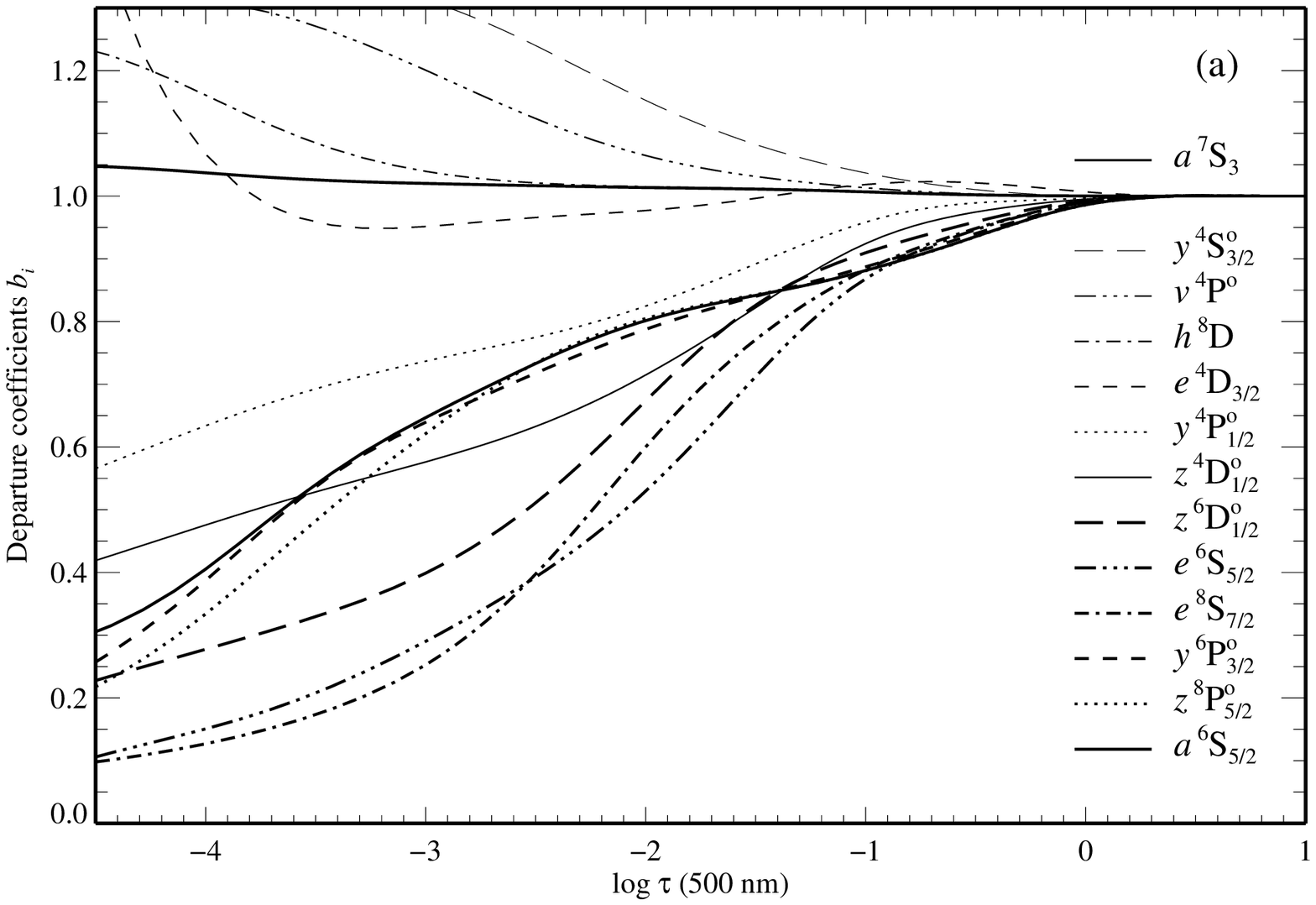}}\hfill
  \resizebox{\columnwidth}{!}{\includegraphics{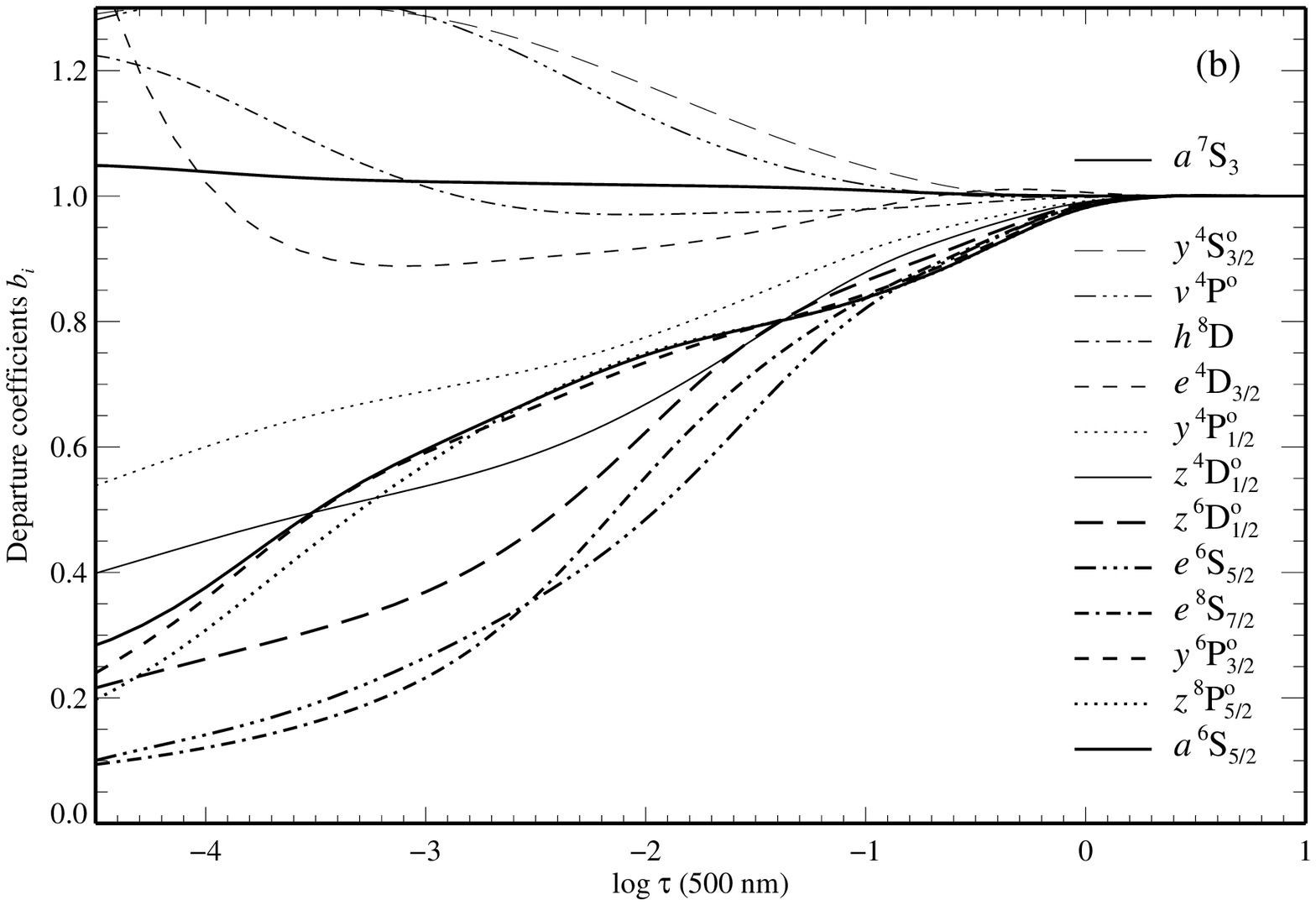}}}
\vspace{0mm}
\hbox{\resizebox{\columnwidth}{!}{\includegraphics{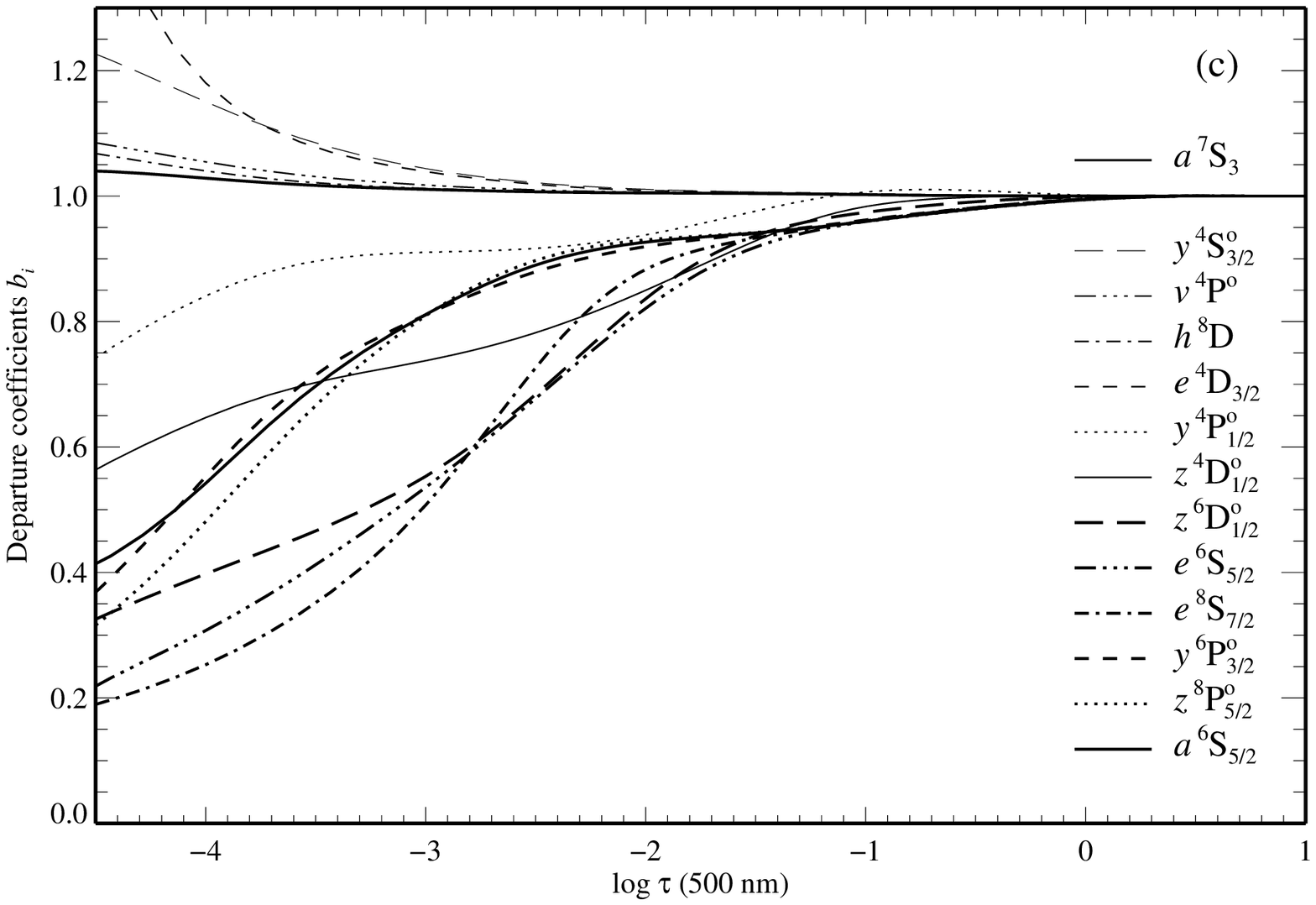}}\hfill
  \resizebox{\columnwidth}{!}{\includegraphics{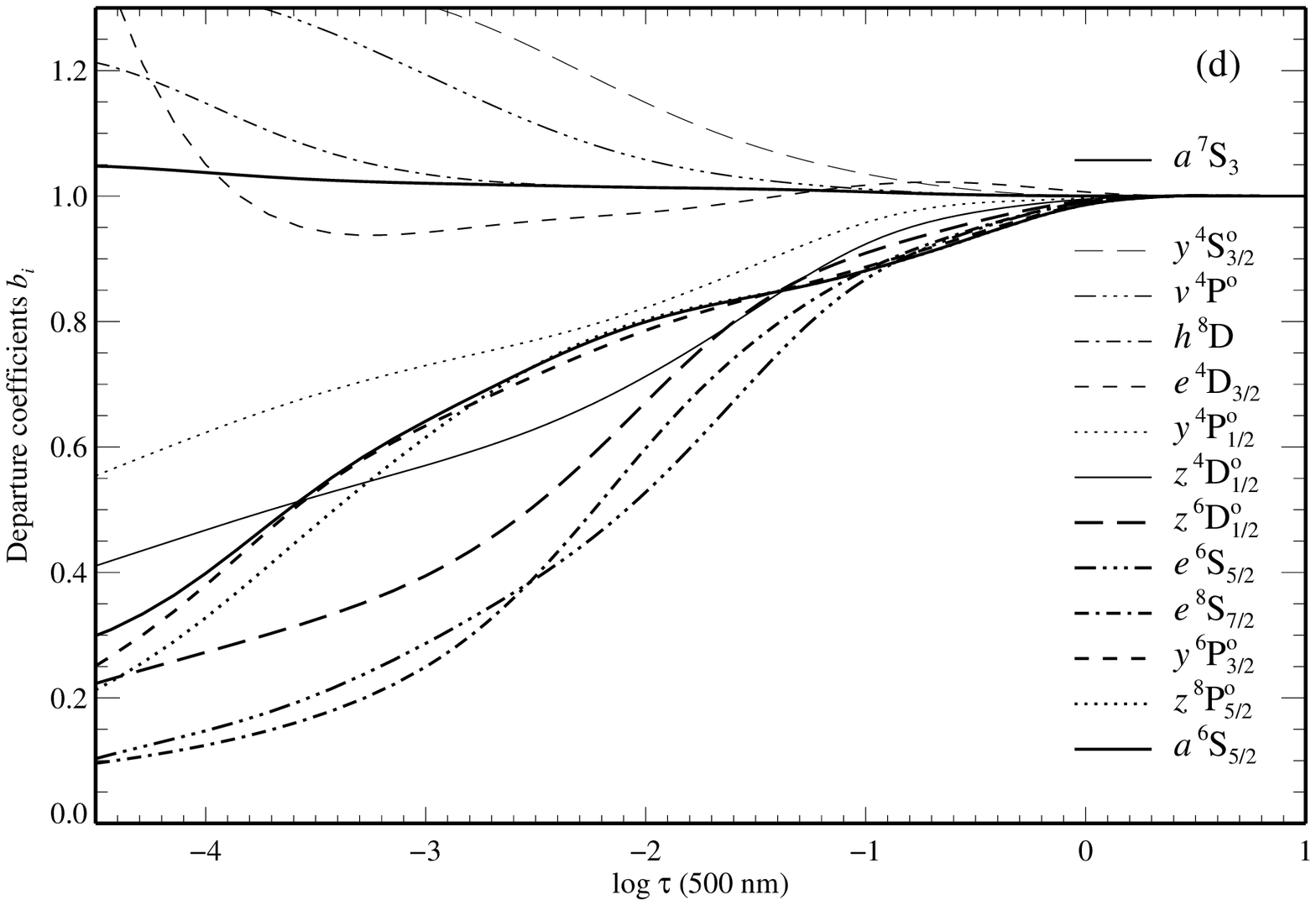}}}
\vspace{0mm}
\hbox{\resizebox{\columnwidth}{!}{\includegraphics{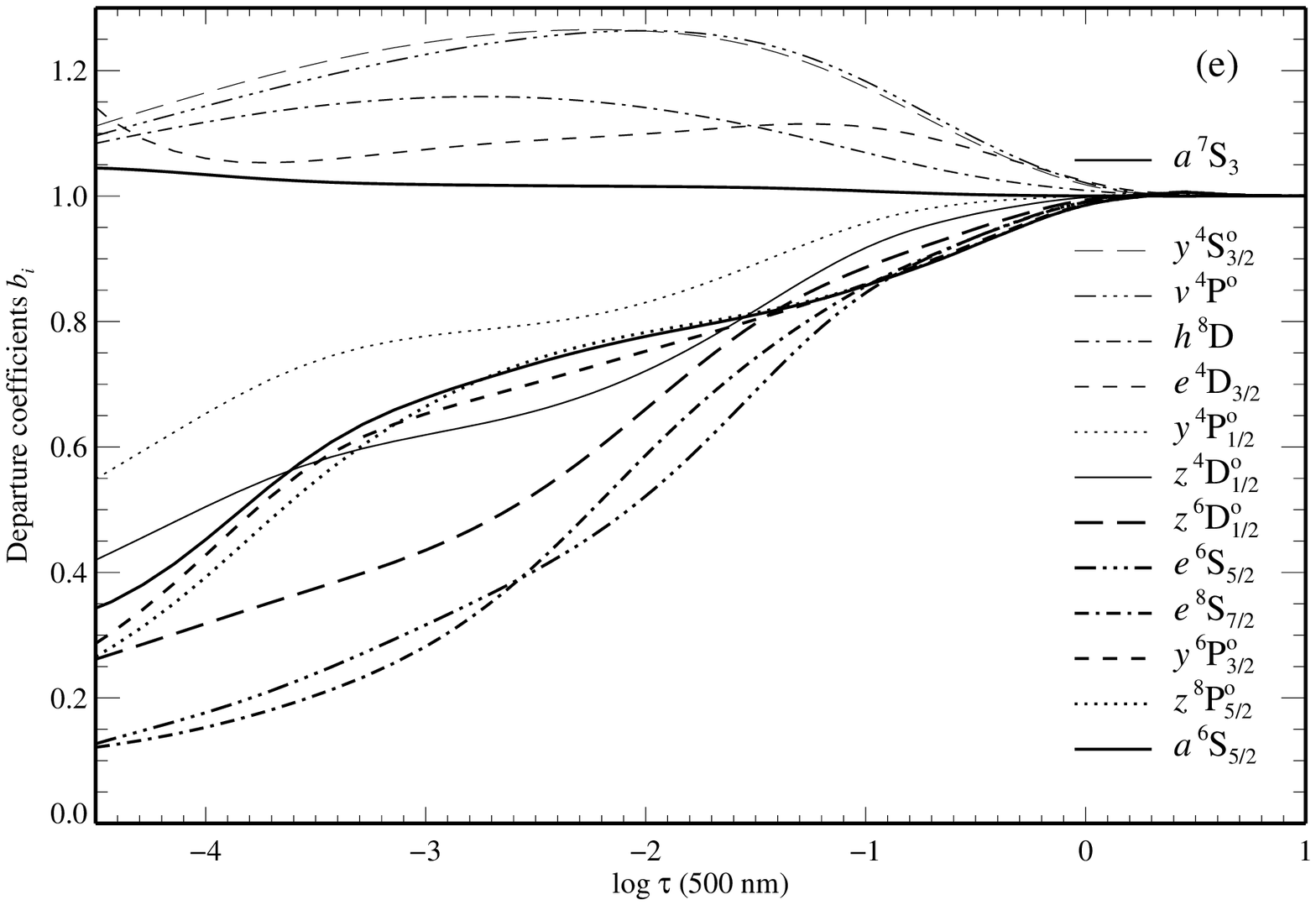}}\hfill
  \resizebox{\columnwidth}{!}{\includegraphics{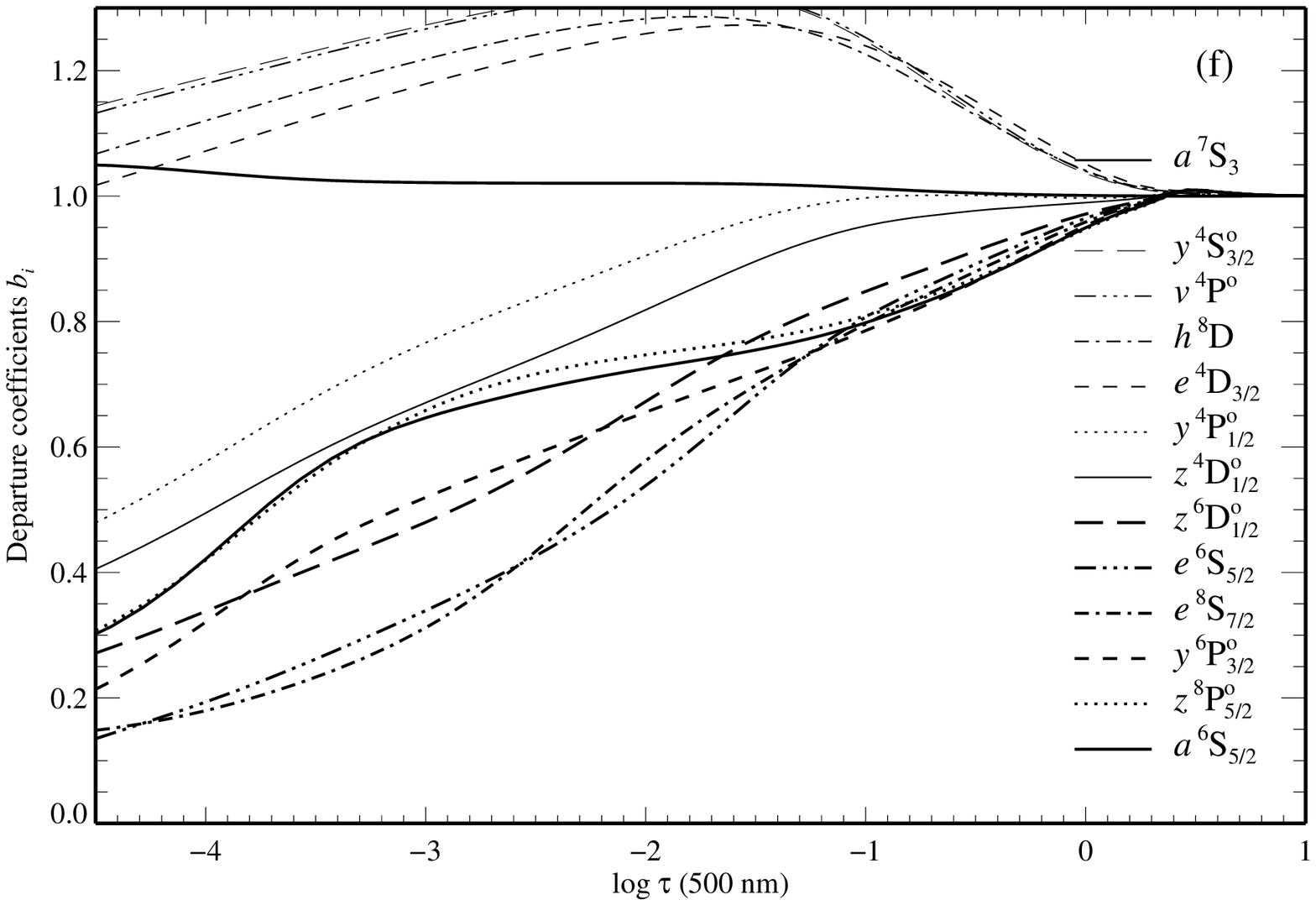}}}
\vspace{0mm} \caption[]{Departure coefficients $b_i$ of selected \ion{Mn}{i}
levels for our standard model atom as a function of optical depth. The curves
are labeled on the right. (a): The reference model atom with a total
of 459 levels. Hydrogen collision rates are scaled by $\SH = 0.05$. The ground
state $\Mn{a}{6}{S}{}{5/2}$ is representative of most of the metastable levels
such as $\Mn{a}{6}{D}{}{}$, $\Mn{a}{4}{D}{}{}$, and $\Mn{a}{4}{G}{}{}$. The odd
levels $\Mn{z}{8}{P}{o}{}$ and $\Mn{z}{6}{P}{o}{}$ are also very similar. Note a
general trend of the departure coefficients increasing with excitation energy.
(b): Same as (a), but hydrogen collisions set equal to zero. The departure
coefficients are very much the same as in the reference model. (c): same as (a),
but $\SH = 5.0$. (d): same as (a), but photoionization rates set equal to
zero.(e): same as (a), but hydrogenic photoionization rates scaled by a factor
300. (f): same as (a), but photoionization rates scaled by a factor 5000.}
\label{bfac} \end{figure*}

Departure coefficients $b_i =  n_i^{\rm NLTE}/n_i^{\rm LTE}$ for
\ion{Mn}{i} levels, calculated under different model assumptions, are presented
in Figs. \ref{bfac} and \ref{reduced}, where Fig. \ref{bfac}a shows the
deviations from LTE for the \emph{reference atom}. Each term is represented by
one level because of the close coupling among the fine-structure levels. In
order not to overload the figure, we have only entered a few selected levels
which are typical for their depth dependence. As an example, the relative
population of the ground state $\Mn{a}{6}{S}{}{5/2}$ is very similar to that of
the other metastable terms, $\Mn{a}{6}{D}{}{}$, $\Mn{a}{4}{D}{}{}$, and
$\Mn{a}{4}{G}{}{}$. The thick curves refer to levels of relatively low
excitation; they are confronted with a number of highly-excited levels that show
a very different behaviour. These levels deviate from the low-excitation levels
following a trend that predicts more thermal or even superthermal populations
the higher the excitation energies are.

In the deepest layers, $b_i = 1$ for all levels (see Fig. \ref{bfac}a).
Deviations from LTE develop between $\log \tau_{\rm 5000} \approx 0.1 \ldots 1$.
At these depths the atmosphere becomes optically thin below the ionization
thresholds of some low-lying levels, such as \Mn{a}{4}{D}{}{} ($\lambda < 2729\ 
\AA$), \Mn{z}{6}{P}{\circ}{} ($\lambda < 2844\ \AA$), \Mn{y}{6}{P}{\circ}{}
($\lambda < 2884\ \AA$), and \Mn{a}{4}{H}{}{} ($\lambda < 3876\ \AA$). The
detailed balance is broken because processes of radiative ionization start
dominating over recombinations, which depend on the local temperature.
Depopulation of these levels is followed by a depopulation of the \ion{Mn}{i}
ground state. This is powered by strong collisional processes, which couple
\Mn{a}{6}{S}{}{5/2} to the excited levels (essentially the metastable ones) and
maintain the relative balance among them out to $\log \tau_{\rm 5000} \approx
-2.8$.

Radiation in the stronger lines also contributes to the redistribution
in the atomic level populations. Line pumping takes action when the optical
depth drops below unity in the wings of lines that belong to multiplets 21, 22
and 23 ( $\lambda\ 4257, 4761, 4739, 4451\ \AA$). A strong non-local UV
radiation field, $J_{\rm \nu} > B_{\nu}(\Te)$, increases the photoexcitation
rates, which are not compensated by the inverse rates of de-excitation. Hence,
at the depths of line formation, below $\log \tau_{\rm 5000} \approx -1.5$,
populations of the upper levels for these transitions (\Mn{y}{4}{P}{\circ}{},
\Mn{z}{4}{D}{\circ}{}, \Mn{z}{4}{F}{\circ}{}) are amplified relative to the
ground state and adjacent low excitation levels. The line source functions
$S_{ij}$ are larger than $B_{\nu}(\Te)$.

The photon suction (Bruls et al. \cite{Bruls92}) effectively operates in the
strong lines of multiplets 27 (transitions from \Mn{e}{6}{S}{}{} to
\Mn{z}{6}{P}{\circ}{}) and 16 (\Mn{e}{8}{S}{}{} to \Mn{z}{8}{P}{\circ}{}). This
process can be understood as a successive radiative de-excitation of an atom
through transitions with high probability in atmospheric layers where
the mean intensity $J_{\rm \nu}$ falls below $B_{\nu}(\Te)$, resulting in
a net downward rate. This leads to a perceptible depopulation of the
upper levels \Mn{e}{8}{S}{}{} and \Mn{e}{6}{S}{}{}, although the corresponding
overpopulation of the lower levels is rather small.

Levels with an excitation energy above 6.5 eV are more closely linked
to the continuum by means of collisions. For the majority of these
high-excitation \ion{Mn}{i} levels the processes of recombination will dominate
instead, which is evident from their negative net ionization rates. The reason
is that at large depths $J_{\rm \nu}$ in the IR spectrum is smaller than
$B_{\nu}(\Te)$. Lack of ionizing photons will result in an increased
net recombination. Hence the levels will be slightly overpopulated,
most of them remaining in detailed balance with \Mn{a}{7}{S}{}{3}.
\subsection{Nature of the NLTE effects}
It is clear from the analysis given in the previous section that there are
several processes that take part in establishing the occupation numbers
of \ion{Mn}{i} and \ion{Mn}{ii} levels.  To identify the major cause of the
non-equilibrium populations, we perform NLTE calculations applying various
scaling factors to photoionization and collision cross-sections. In addition, we
investigate how the completeness of the model atom affects the results.
\subsubsection{Standard models}
\begin{figure*}
\hbox{\resizebox{\columnwidth}{!}{\includegraphics{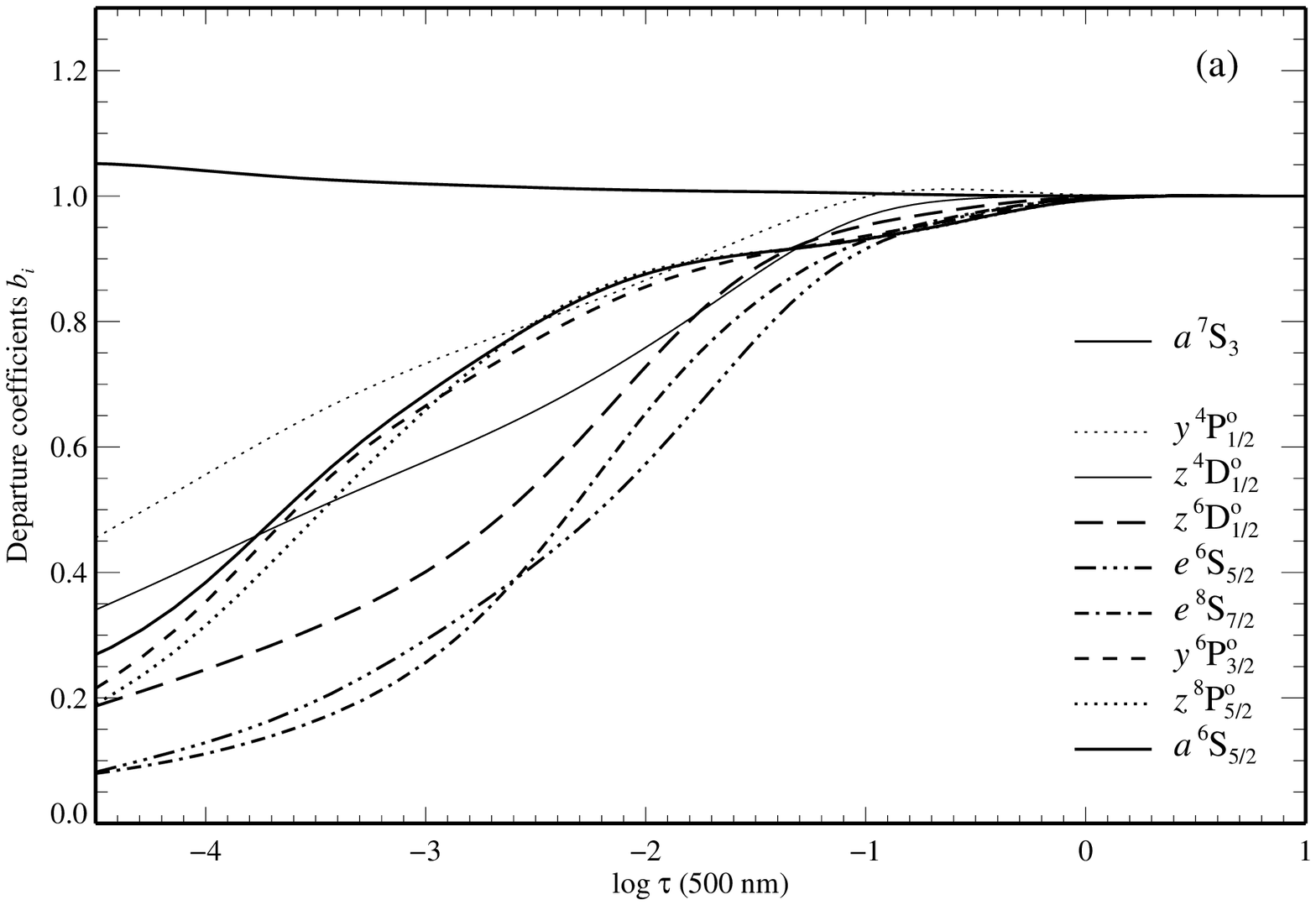}}\hfill
  \resizebox{\columnwidth}{!}{\includegraphics{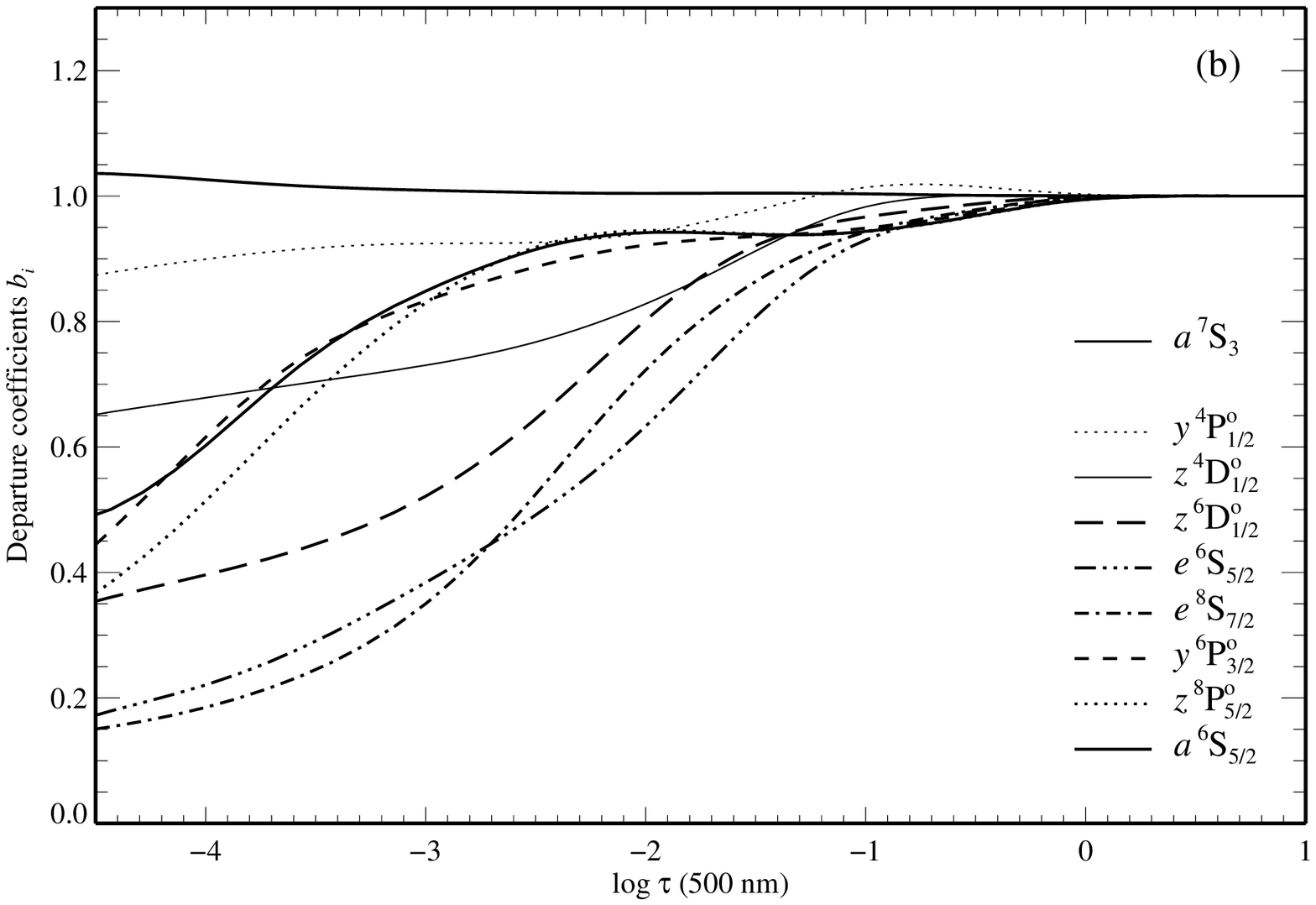}}}
\vspace{0mm}
\hbox{\resizebox{\columnwidth}{!}{\includegraphics{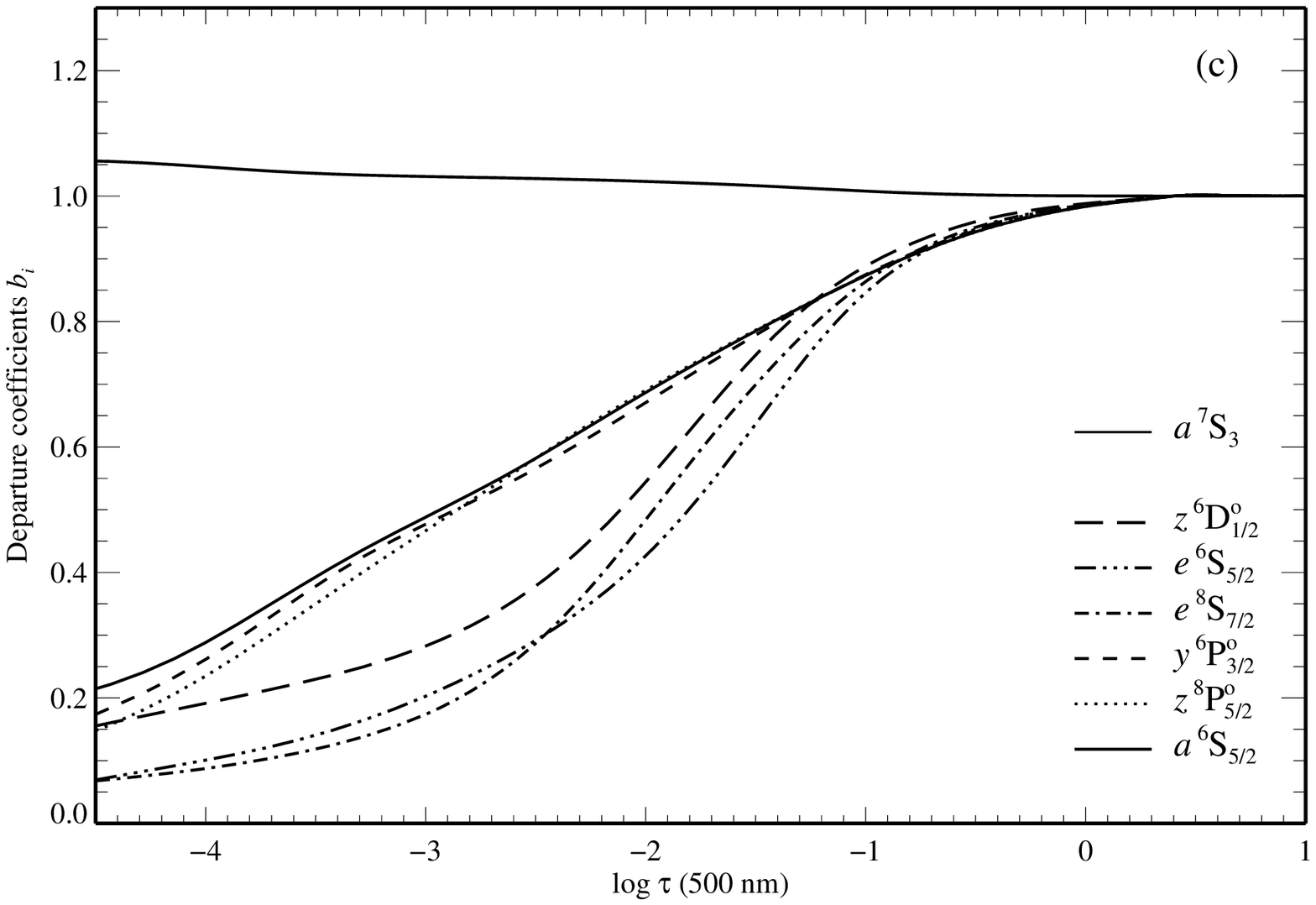}}\hfill
  \resizebox{\columnwidth}{!}{\includegraphics{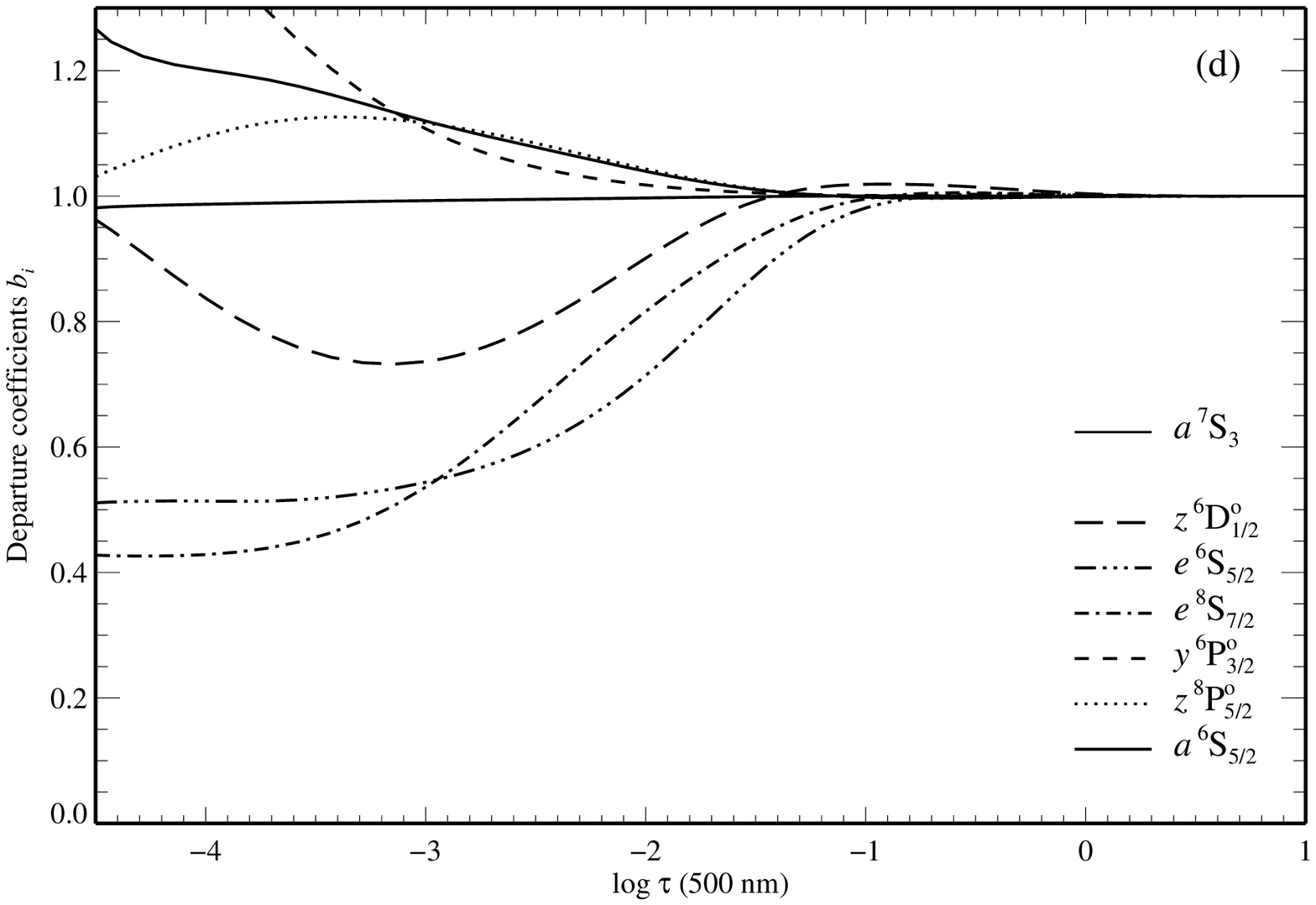}}}
\vspace{0mm}
\hbox{\resizebox{\columnwidth}{!}{\includegraphics{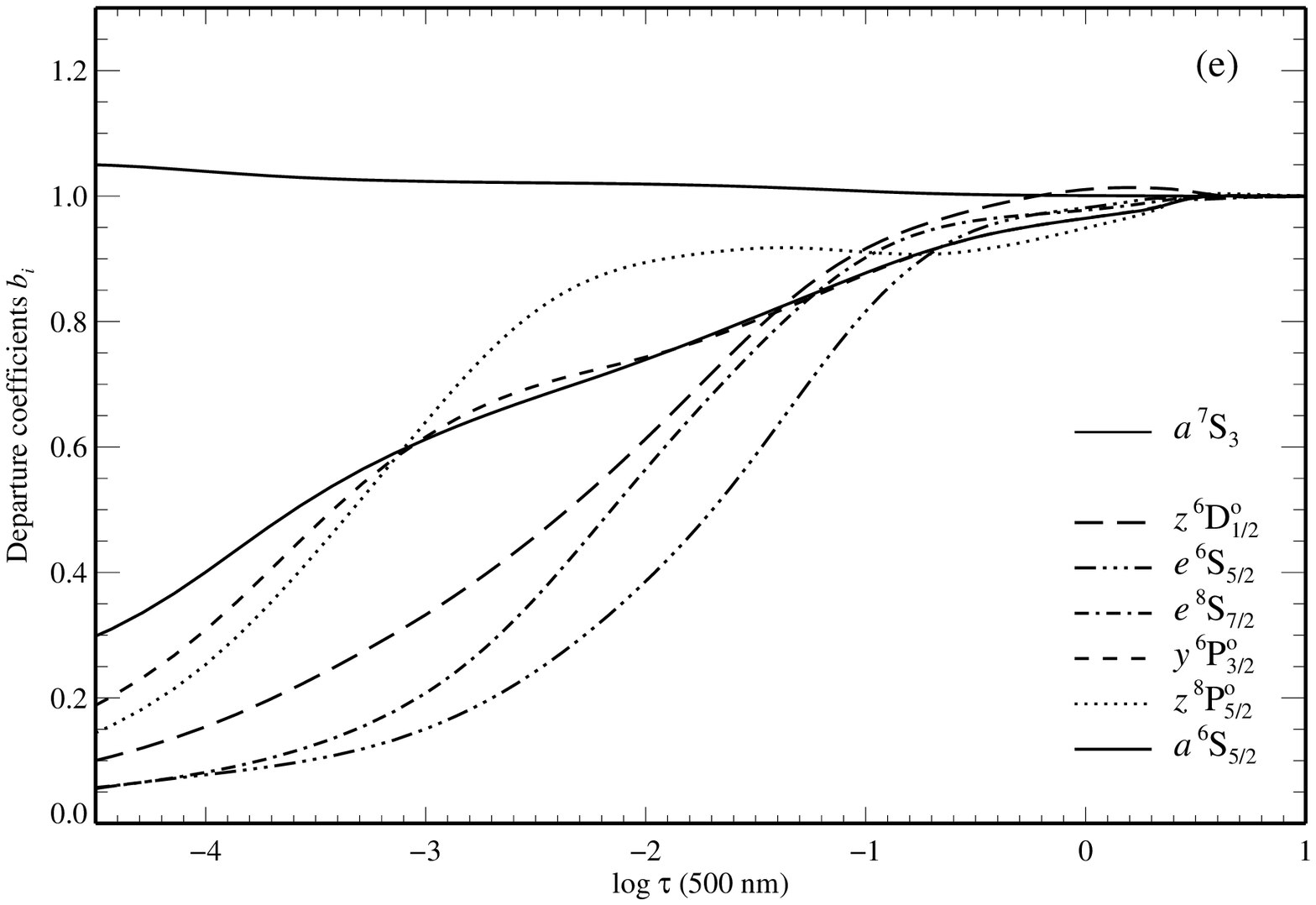}}\hfill
  \resizebox{\columnwidth}{!}{\includegraphics{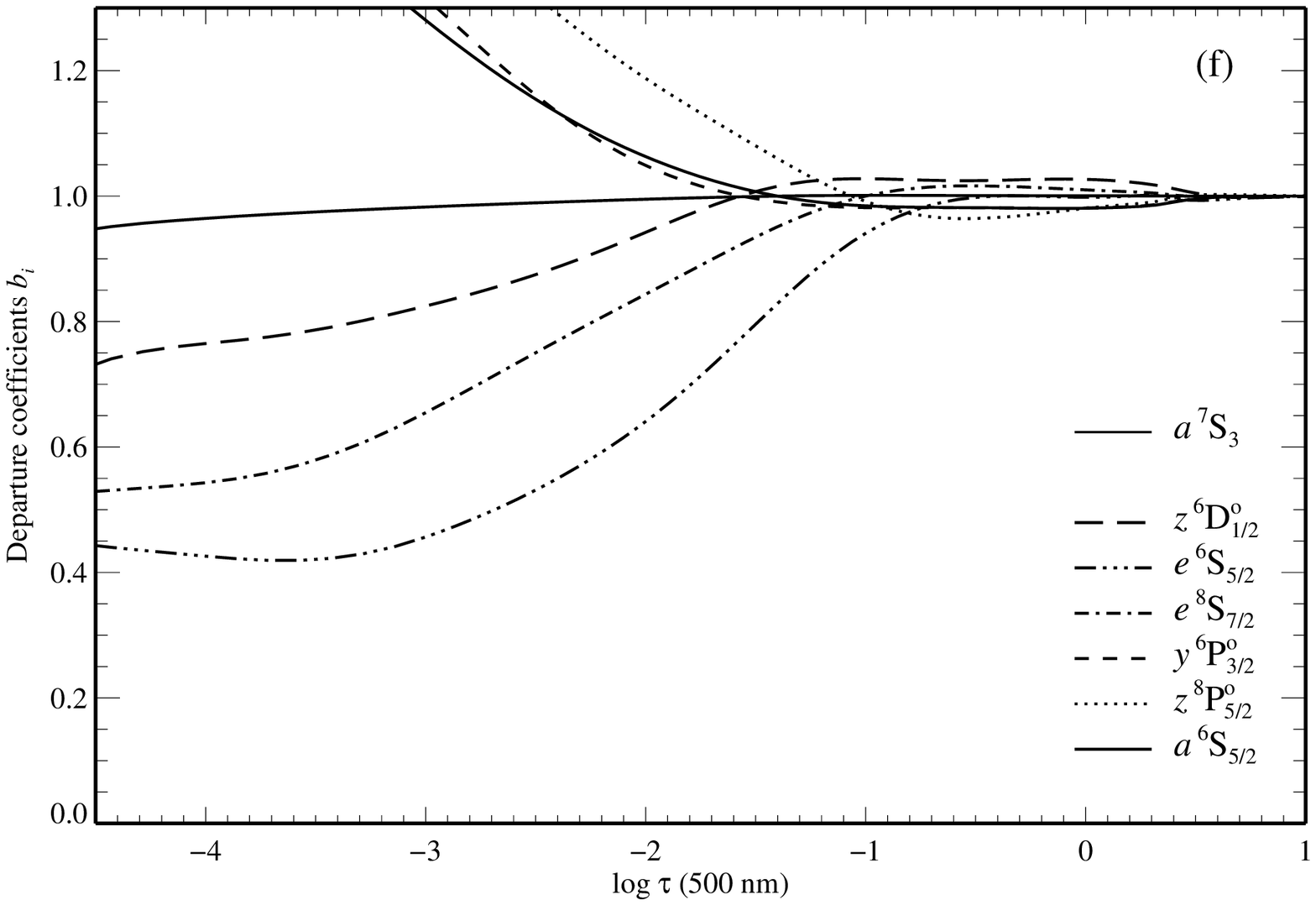}}}
\vspace{0mm} \caption[]{Departure coefficients $b_i$ of selected \ion{Mn}{i}
levels for reduced model atoms as a function of optical depth. The curves are
labeled on the right. (a): The reduced model atom with a total of 146
levels including the \ion{Mn}{ii} ground state. Hydrogen collision rates are
scaled by $\SH = 0.05$ as in the reference atom. Note that the uppermost levels
of Fig. \ref{bfac} are no longer represented by the reduced model. (b):
Same as (a), but photoionization rates set equal to zero. (c): The
reduced model atom with a total of 66 levels including the \ion{Mn}{ii} ground
state. Hydrogen collision rates are scaled by $\SH = 0.05$ as in the reference
atom. (d): same as (c), but photoionization rates set equal to zero.
(e): same as (c), but electron collision rates set equal to zero.
(f): same as (d), but electron collision rates set equal to zero.}
\label{reduced}
\end{figure*}
The reference atomic model of \ion{Mn}{i}, as described in Sect. \ref{refmod}, 
was constructed with hydrogenic photoionization cross-sections and
cross-sections for inelastic collisions with \ion{H}{i} (both for b-b and b-f
transitions) based on Drawin's formula, but scaled by a factor $\SH = 0.05$.
Comparison of the collision rates $C_{ij}$(H) and $C_{ij}$(e) shows that in the
solar atmosphere for \ion{Mn}{i} rates of collisions with electrons are on
average greater by one order of magnitude. In order to test the influence of
inelastic collisions with hydrogen, we carried out additional calculations for
scaling factors from $\SH = 0$ to $5$ (Fig. \ref{bfac}a-c). Naturally, with
increasing rates of collisional excitations and ionizations by \ion{H}{i} atoms
departures get closer to unity, although perfect thermalization is not achieved
with values for $\SH$ as low as 5.

The influence of photoionization was studied likewise, i.e., by means of
performing NLTE calculations for various scaling factors to the cross-sections.
For test purposes we have chosen to scale the \ion{Mn}{i} cross-sections
$\sigma_\nu$ by factors $\SP = 0$, 300 and 5000. A plot of the
resulting departure coefficients is presented in Fig. \ref{bfac}d-f. Absence of
photoionization (Fig. \ref{bfac}d) does not restore LTE, as we could expect if
the major and only NLTE mechanism were the overionization. Moreover, the
results for zero photoionization cross-sections are virtually indistinguishable
from those of the reference model ($\SP = 1$), while $\SP = 300$ leads to
slightly stronger deviations of $b_i$ from unity (Fig. \ref{bfac}e). When
$\sigma_\nu$ are increased by a factor of 5000 (Fig. \ref{bfac}f) the
overionization is amplified and all \ion{Mn}{i} levels are noticeably affected.
\subsubsection{Reduced models}
We consider reduced model atoms that are discussed in detail in this section. In
Fig. \ref{reduced} we show departure coefficients for two reduced models, one
with 145 levels of \ion{Mn}{i} and an excitation potential of the uppermost
level of 6.42 eV ($E_{\rm ion} = 7.43$ eV), and one with 65 \ion{Mn}{i} levels
limited to 5.23 eV. For both models only the \ion{Mn}{ii} ground state was
considered. Note that due to the model reduction some of the high-excitation
levels in Fig. \ref{bfac} are no longer present.

Fig. \ref{reduced}a and b display the departure coefficients for the
146 level atom. As a result of a reduced recombination efficiency the remaining
high-excitation levels (not shown here) are slightly less overpopulated, when
compared to the NLTE populations calculated with the full atomic model. Fig.
\ref{reduced}b shows $b_i$ calculated with zero photoionization. Most of the
low-excitation levels are still \emph{underpopulated} due to pumping processes
combined with an efficient collisional coupling to the \ion{Mn}{ii} ground
state. It is only for the 66 level model atom that the domination of
photoionization can be seen. Fig. \ref{reduced}c and d demonstrate the loss of
the hydrogen collision efficiency in coupling to the \ion{Mn}{ii} ground state.
Here, removal of photoionization with $\SP = 0$ drives the lower level
populations towards LTE, at least in the line formation region. In the absence
of photoionization the photon suction along the lines now results in an
\emph{overpopulation} of low-excitation \ion{Mn}{i} levels that is balanced by
net collisional ionization. Neglecting electron collisions (Fig. \ref{reduced}e
and f) results in an even weaker coupling of the levels.

\begin{table*}
\begin{minipage}{\linewidth}
\renewcommand{\footnoterule}{}  
\caption{Parameters of lines used for abundance calculation. An asterisk (*) in
the wavelength entry refers to the lines with consistent ocillator strengths
from different sources (see text). No error estimate is given for Kurucz $\log
gf $ values.} \label{lines} 
\begin{center}
\begin{tabular}{rl|rrlcc|rrl|rlll}
\noalign{\smallskip}\hline\noalign{\smallskip} No. & ~~~~~~$\lambda$ & Mult. &
$N_{\rm HFS}$ & $\Elow$ & Lower & Upper &
$W_\lambda$ & $\log gf $ & Error & $\Delta \log \varepsilon$~~~ & $\Delta X$ &
$\loggfe$ & Source\footnote{References: ~~~(1) Booth et al. \cite{Booth84a}; (2)
Greenlee \& Whaling \cite{Green79}; (3) Blackwell \& Collins \cite{Black72}; (4)
Woodgate \cite{Wood66}; (5) Kurucz \cite{Kurucz88}}\\
     & ~~~~[\AA] &  &  & [eV] & level & level & [\mA] &   & & &  &  & \\
\noalign{\smallskip}\hline\noalign{\smallskip}
1  & 4055.513* &  5  & 4 &  2.13  &  \Mn{a}{6}{D}{ }{7/2} & \Mn{z}{6}{D}{o}{7/2}
& 136.~~  & --0.070 &  0.1  & --0.15~~  &~~0.0  & 5.250 &  1\\
2  & 4058.911* &  5  & 3 &  2.17  &  \Mn{a}{6}{D}{ }{3/2} & \Mn{z}{6}{D}{o}{1/2}
& 101.~~  & --0.446 &  0.1  & --0.05~~  &~~0.0  & 4.974 &  1\\
3  & 4070.264  &  5  & 3 &  2.19  &  \Mn{a}{6}{D}{ }{1/2} & \Mn{z}{6}{D}{o}{1/2}
& 70.~~   & --0.950 &  0.1  & --0.06~~  &~~0.06 & 4.460 &  3\\
4  & 4257.641  &  23 & 3 &  2.94  &  \Mn{a}{4}{D}{ }{1/2} & \Mn{y}{4}{P}{o}{1/2}
& 62.~~   & --0.700 &  0.18 &   0.11~~  &~~0.095& 4.880 &  4\\
5  & 4436.342  &  22 & 3 &  2.91  &  \Mn{a}{4}{D}{ }{5/2} & \Mn{z}{4}{D}{o}{3/2}
& 71.3    & --0.288 &  0.04 & --0.21~~  &~~0.07 & 4.972 &  1\\
6  & 4451.581  &  22 & 3 &  2.88  &  \Mn{a}{4}{D}{ }{7/2} & \Mn{z}{4}{D}{o}{7/2}
& 93.~~   &   0.278 &  0.04 & --0.3~~~~ &~~0.0  & 5.448 &  1\\
7  & 4453.001  &  22 & 2 &  2.93  &  \Mn{a}{4}{D}{ }{3/2} & \Mn{z}{4}{D}{o}{1/2}
& 53.5    & --0.490 &  0.04 & --0.165   &~~0.08 & 4.815 &  1\\
8  & 4455.288  &  28 & 3 &  3.07  &  \Mn{z}{6}{P}{o}{3/2} & \Mn{e}{6}{D}{ }{3/2}
& 73.~~   & --0.246 &       &   0.00~~  &~~0.05 & 5.224 &  5\\
9  & 4457.010* &  28 & 3 &  3.06  &  \Mn{z}{6}{P}{o}{5/2} & \Mn{e}{6}{D}{ }{3/2}
& 50.~~   & --0.555 &  0.1  & --0.185   &~~0.06 & 4.730 &  1\\
10 & 4470.143  &  22 & 2 &  2.93  &  \Mn{a}{4}{D}{ }{3/2} & \Mn{z}{4}{D}{o}{3/2}
& 53.5    & --0.444 &  0.04 & --0.23~~  &~~0.07 & 4.796 &  1\\
11 & 4490.067  &  22 & 2 &  2.94  &  \Mn{a}{4}{D}{ }{1/2} & \Mn{z}{4}{D}{o}{3/2}
& 56.~~   & --0.522 &  0.04 & --0.16~~  &~~0.065& 4.788 &  1\\
12 & 4498.901  &  22 & 2 &  2.93  &  \Mn{a}{4}{D}{ }{3/2} & \Mn{z}{4}{D}{o}{5/2}
& 57.~~   & --0.343 &  0.04 & --0.37~~  &~~0.06 & 4.757 &  1\\
13 & 4502.220  &  22 & 2 &  2.91  &  \Mn{a}{4}{D}{ }{5/2} & \Mn{z}{4}{D}{o}{7/2}
& 59.~~   & --0.345 &  0.04 & --0.37~~  &~~0.05 & 4.755 &  1\\
14 & 4671.667* &  21 & 5 &  2.88  &  \Mn{a}{4}{D}{ }{7/2} & \Mn{z}{4}{F}{o}{5/2}
& 12.8    & --1.670 &  0.1  & --0.13~~  &~~0.08 & 3.670 &  2\\
15 & 4709.705  &  21 & 4 &  2.88  &  \Mn{a}{4}{D}{ }{7/2} & \Mn{z}{4}{F}{o}{7/2}
& 72.~~   & --0.340 &  0.04 & --0.35~~  &~~0.06 & 4.780 &  1\\
16 & 4739.088  &  21 & 4 &  2.93  &  \Mn{a}{4}{D}{ }{3/2} & \Mn{z}{4}{F}{o}{3/2}
& 62.~~   & --0.490 &  0.04 & --0.24~~  &~~0.07 & 4.740 &  1\\
17 & 4754.021* &  16 & 5 &  2.27  &  \Mn{z}{8}{P}{o}{5/2} & \Mn{e}{8}{S}{ }{7/2}
& 146.~~  & --0.086 &  0.04 & --0.16~~  &--0.04 & 5.224 &  1\\
18 & 4761.508  &  21 & 4 &  2.94  &  \Mn{a}{4}{D}{ }{1/2} & \Mn{z}{4}{F}{o}{3/2}
& 73.~~   & --0.138 &  0.04 & --0.18~~  &~~0.07 & 5.152 &  1\\
19 & 4762.358  &  21 & 5 &  2.88  &  \Mn{a}{4}{D}{ }{7/2} & \Mn{z}{4}{F}{o}{9/2}
& 108.~~  &   0.425 &  0.04 & --0.36~~  &~~0.0  & 5.535 &  1\\
20 & 4765.851  &  21 & 3 &  2.93  &  \Mn{a}{4}{D}{ }{3/2} & \Mn{z}{4}{F}{o}{5/2}
& 81.~~   & --0.080 &  0.1  & --0.17~~  &~~0.05 & 5.220 &  1\\
21 & 4766.413  &  21 & 4 &  2.91  &  \Mn{a}{4}{D}{ }{5/2} & \Mn{z}{4}{F}{o}{7/2}
& 98.5    &   0.100 &  0.1  & --0.23~~  &~~0.0  & 5.340 &  2\\
22 & 4783.389* &  16 & 5 &  2.29  &  \Mn{z}{8}{P}{o}{7/2} & \Mn{e}{8}{S}{ }{7/2}
& 148.~~  &   0.042 &  0.04 & --0.32~~  &~~0.0  & 5.192 &  1\\
23 & 4823.460* &  16 & 6 &  2.31  &  \Mn{z}{8}{P}{o}{9/2} & \Mn{e}{8}{S}{ }{7/2}
& 149.~~  &   0.144 &  0.04 & --0.32~~  &--0.03 & 5.294 &  1\\
24 & 5004.894  &  20 & 4 &  2.91  &  \Mn{a}{4}{D}{ }{5/2} & \Mn{z}{6}{F}{o}{7/2}
& 13.7    & --1.630 &  0.1  & --0.12~~  &~~0.08 & 3.720 &  2\\
25 & 5117.913* &  32 & 3 &  3.12  &  \Mn{a}{4}{G}{ }{5/2} & \Mn{z}{4}{F}{o}{3/2}
& 24.2    & --1.140 &  0.1  & --0.05~~  &~~0.08 & 4.280 &  2\\
26 & 5255.287* &  32 & 6 &  3.12  &  \Mn{a}{4}{G}{ }{11/2}& \Mn{z}{4}{F}{o}{9/2}
& 41.5    & --0.763 &  0.04 & --0.2~~~~ &~~0.08 & 4.507 &  1\\
27 & 5394.619* &  1  & 6 &  0     &  \Mn{a}{6}{S}{ }{5/2} & \Mn{z}{8}{P}{o}{7/2}
& 79.5    & --3.503 &  0.1  & --0.115   &~~0.105& 1.852 &  1\\
28 & 5407.331* &  4  & 10 & 2.13  &  \Mn{a}{6}{D}{ }{7/2} & \Mn{y}{6}{P}{o}{7/2}
& 53.~~   & --1.743 &  0.1  & --0.125   &~~0.085& 3.602 &  1\\
29 & 5420.265* &  4  & 9 &  2.13  &  \Mn{a}{6}{D}{ }{7/2} & \Mn{y}{6}{P}{o}{5/2}
& 85.~~   & --1.462 &  0.1  & --0.06~~  &~~0.09 & 3.948 &  1\\
30 & 5432.512* &  1  & 5 &  0     &  \Mn{a}{6}{S}{ }{5/2} & \Mn{z}{8}{P}{o}{5/2}
& 48.~~   & --3.795 &  0.1  & --0.12~~  &~~0.11 & 1.555 &  1\\
31 & 5470.560  &  4  & 8 &  2.15  &  \Mn{a}{6}{D}{ }{5/2} & \Mn{y}{6}{P}{o}{5/2}
& 57.5    & --1.702 &       & --0.04~~  &~~0.085& 3.728 &  5\\
32 & 5516.697* &  4  & 8 &  2.17  &  \Mn{a}{6}{D}{ }{3/2} & \Mn{y}{6}{P}{o}{3/2}
& 44.~~   & --1.847 &  0.1  & --0.02~~  &~~0.085& 3.603 &  1\\
33 & 5537.692  &  4  & 5 &  2.18  &  \Mn{a}{6}{D}{ }{1/2} & \Mn{y}{6}{P}{o}{3/2}
& 36.~~   & --2.017 &       & --0.02~~  &~~0.085& 3.433 &  5\\
34 & 6013.465  &  27 & 6 &  3.06  &  \Mn{z}{6}{P}{o}{3/2} & \Mn{e}{6}{S}{ }{5/2}
& 87.~~   & --0.251 &  0.1  & --0.285   &--0.015& 4.934 &  1\\
35 & 6016.586  &  27 & 6 &  3.06  &  \Mn{z}{6}{P}{o}{5/2} & \Mn{e}{6}{S}{ }{5/2}
& 97.8    & --0.216 &       & --0.145   &--0.05 & 5.109 &  5\\
36 & 6021.727  &  27 & 6 &  3.06  &  \Mn{z}{6}{P}{o}{7/2} & \Mn{e}{6}{S}{ }{5/2}
& 96.8    &   0.034 &  0.1  & --0.27~~  &--0.07 & 5.234 &  1\\
37 & 8700.877  &  49 & 10 & 4.41  &  \Mn{y}{6}{P}{o}{5/2} & \Mn{e}{6}{D}{ }{5/2}
& 11.~~   & --0.477 &       & --0.058   &~~0.1  & 4.935 &  5\\
38 & 8703.559  &  49 & 11 & 4.41  &  \Mn{y}{6}{P}{o}{5/2} & \Mn{e}{6}{D}{ }{7/2}
& 15.5    & --0.328 &       & --0.057   &~~0.1  & 5.085 &  5\\
39 & 8740.705  &  49 & 13 & 4.42  &  \Mn{y}{6}{P}{o}{7/2} & \Mn{e}{6}{D}{ }{9/2}
& 26.~~   & --0.055 &       & --0.09~~  &~~0.09 & 5.325 &  5\\
\noalign{\smallskip}\hline\noalign{\smallskip}
\end{tabular}
\end{center}
\end{minipage}
\end{table*}
\subsubsection{Discussion}
The behaviour of Mn level populations, strongly deviating from LTE, can
not be explained simply in terms of photoionization or collisions, as was
succesfully done for other elements submitted to an NLTE analysis. Our results
show that hydrogenic b-f radiative cross-sections for the \ion{Mn}{i} levels of
3 to 4 eV excitation are too small to ensure the dominance of photoionization,
which is the main NLTE effect for many metal-group elements. These levels
usually dominate the ionization balance in such ions because in the atmospheres
of solar-type  stars their photoionization edges lie in regions where the
radiation field is intense enough and the radiative rates prevail over
collisional rates. For instance, the Opacity Project calculations (Bautista
\cite{Baut97}) give evidence for neutral iron cross-sections, which are
10 to 1000 times larger than the hydrogenic ones. This could explain the nearly
identical behaviour of Mn departure coefficents for $\SP = 0$ and $\SP = 1$.
Kramers' approximation for \ion{Mn}{i} gives a \emph{lower limit} for the
cross-sections, which is comparable in its influence on populations to
\emph{the absence of photoionization}. Thus, a very large enhancement to
$\sigma_\nu$ is required in order to force the dominance of this process in
establishing the occupation numbers of \ion{Mn}{i} and \ion{Mn}{ii} levels.

The complexity of the \ion{Mn}{i} atomic system with its large number of
levels produces some additional complications for NLTE calculations. There are
many doubly excited (DE) configurations, which, in contrast to those of single
excitation (SE), can not be treated in hydrogenic approximation. The use of
Kramers' formula with principal quantum number $n$ for all atomic levels is an
obvious simplification: DE and SE configurations of neighbouring excitation
energies but different $n$ will have strongly differing photoionization
cross-sections, which scale with inverse fifth power of the principal quantum
number. Furthermore, the b-f collision cross-sections, calculated using the
radiative b-f cross-sections, will differ for these neighbouring levels by some
orders of magnitude. The latter is important for the high-excitation levels 
whose b-f collision rates compensate radiative b-f rates from the low-excitation
levels and in this way establish the statistical equilibrium of the model atom.
As seen on Fig. \ref{reduced}c and d exclusion of the upper levels with $\sim $
1 -- 2 eV energy separation from the continuum, leads to a reduced recombination
efficiency, and the \ion{Mn}{i} atom could be identified as a member of the
photoionization-dominated category.

One of the possible approaches is to use the \emph{effective} principal
quantum numbers of the levels. We have performed NLTE calculations for the
photoionization cross-sections computed from Kramers' formula with $n_{\rm
eff}$. The switch from principal to effective quantum numbers did not change the
results significantly except that the uppermost levels are now coupled more
tightly to the \ion{Mn}{ii} ground state at the line formation depths. Also, the
whole atomic system became more sensitive to the enhancement of photoionization
cross-sections, but levels that produce the lines of interest in our research
were not affected. This solution may be more appropriate for very complex atomic
systems, such as Mn, but at the same time it emphasizes the \emph{inadequacy of
the hydrogenic approximation}. This is, however, nothing new; it was also shown
by Nahar \& Pradhan (\cite{Nahar04}) that photoionization of excited levels of a
non-hydrogenic ion may not be treated in hydrogenic approximation. As there is
currently no alternative to the latter, we decided to preserve the hydrogenic
photoionization cross-sections.

The incompleteness of the available data for the transitions in the Mn
model atom presents another problem. There are more than 500 experimental levels
reported for \ion{Mn}{i}, 180 being above the lowest ionization limit (Johansson
\& Cowley \cite{Johan88}). Although the system of atomic levels seems to be
complete, the case is somewhat different for the allowed transitions between
them. In particular, there is no data in Kurucz's database (Kurucz \& Bell
\cite{Kurucz95}), which we use in our analysis, for transitions that link the
uppermost levels and levels of intermediate energies. Unfortunately, it is
exactly these transitions that should couple the bulk of the overpopulated
levels to the remaining underpopulated levels in our model.

In order to check the results obtained with the DETAIL code, we performed
statistical equilibrium calculations for Mn using the independent NONLTE3
code, based on the method of complete linearization (Sakhibullin \cite{Sakh83}).
The NONLTE3 atomic level populations are in close agreement with those from the 
DETAIL code. Hence, we do not associate the behaviour of the Mn atomic system
with program errors. We suggest that the \ion{Mn}{i} ion represents a complex
\emph{mixed-domination} case, where inelastic collisions with electrons and
hydrogen atoms counteract photoionization and decouple the atomic system.
\section{Synthesis of \ion{Mn}{i} lines}
Spectrum synthesis is employed to determine the abundance of Mn in the
solar atmosphere. We use our line analysis code SIU, which generates synthetic
NLTE line profiles for computed level departure coefficients $b_i(\tau)$. For
all elements, except Mn, we assume LTE.
\begin{figure*}
\hbox{\resizebox{\columnwidth}{!}{\includegraphics{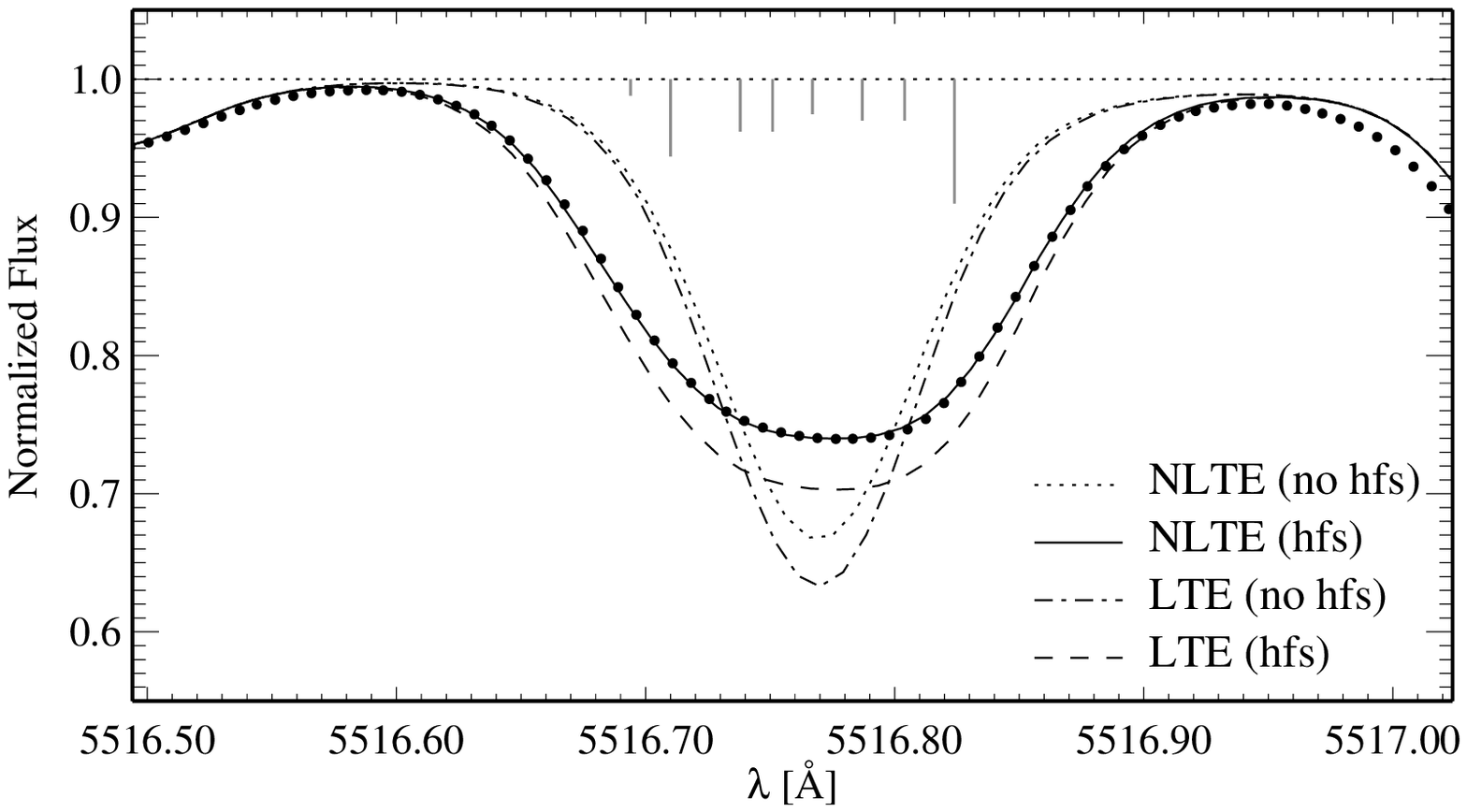}} \hfill
      \resizebox{\columnwidth}{!}{\includegraphics{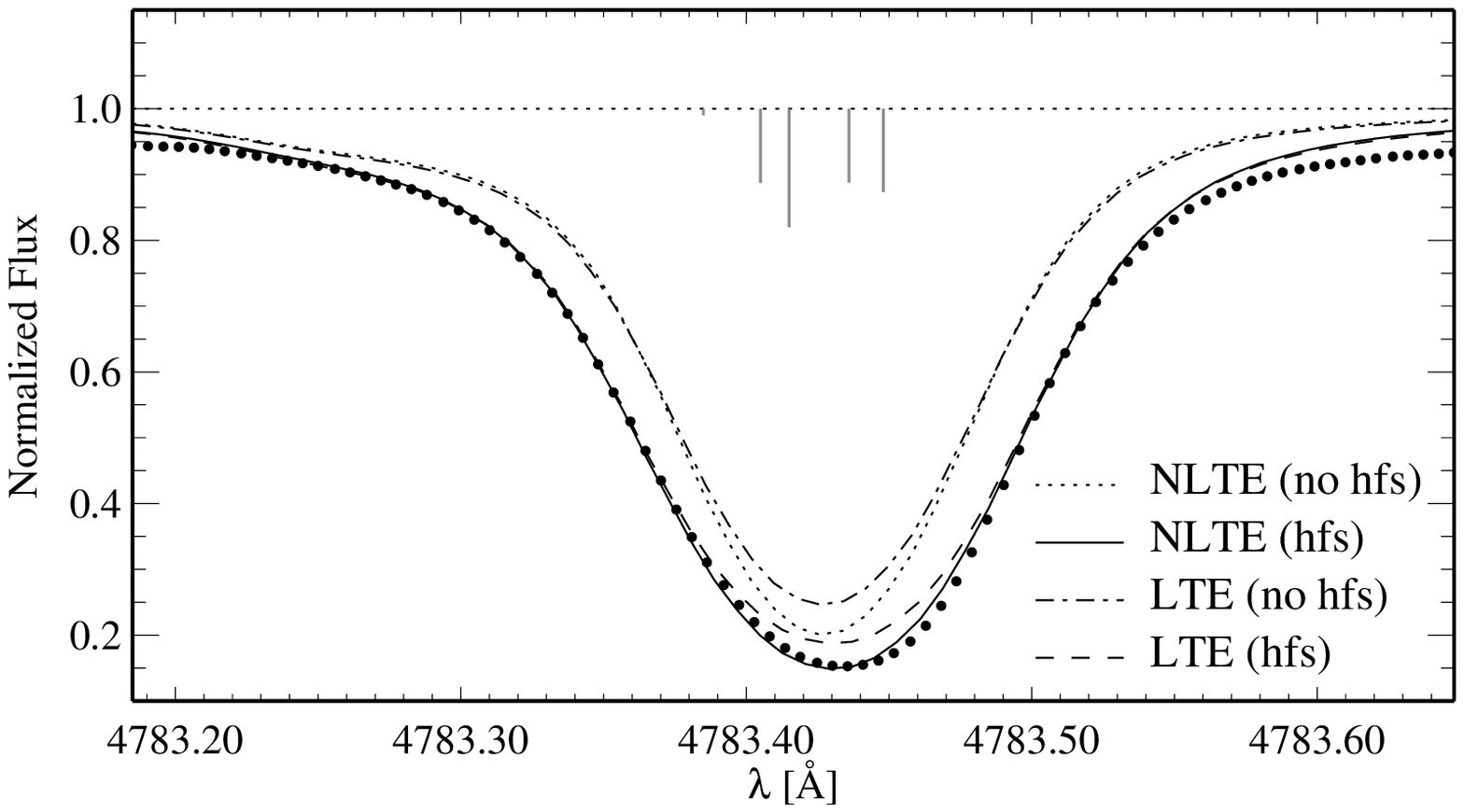}}}
\vspace{-4mm}
\hbox{\resizebox{\columnwidth}{!}{\includegraphics{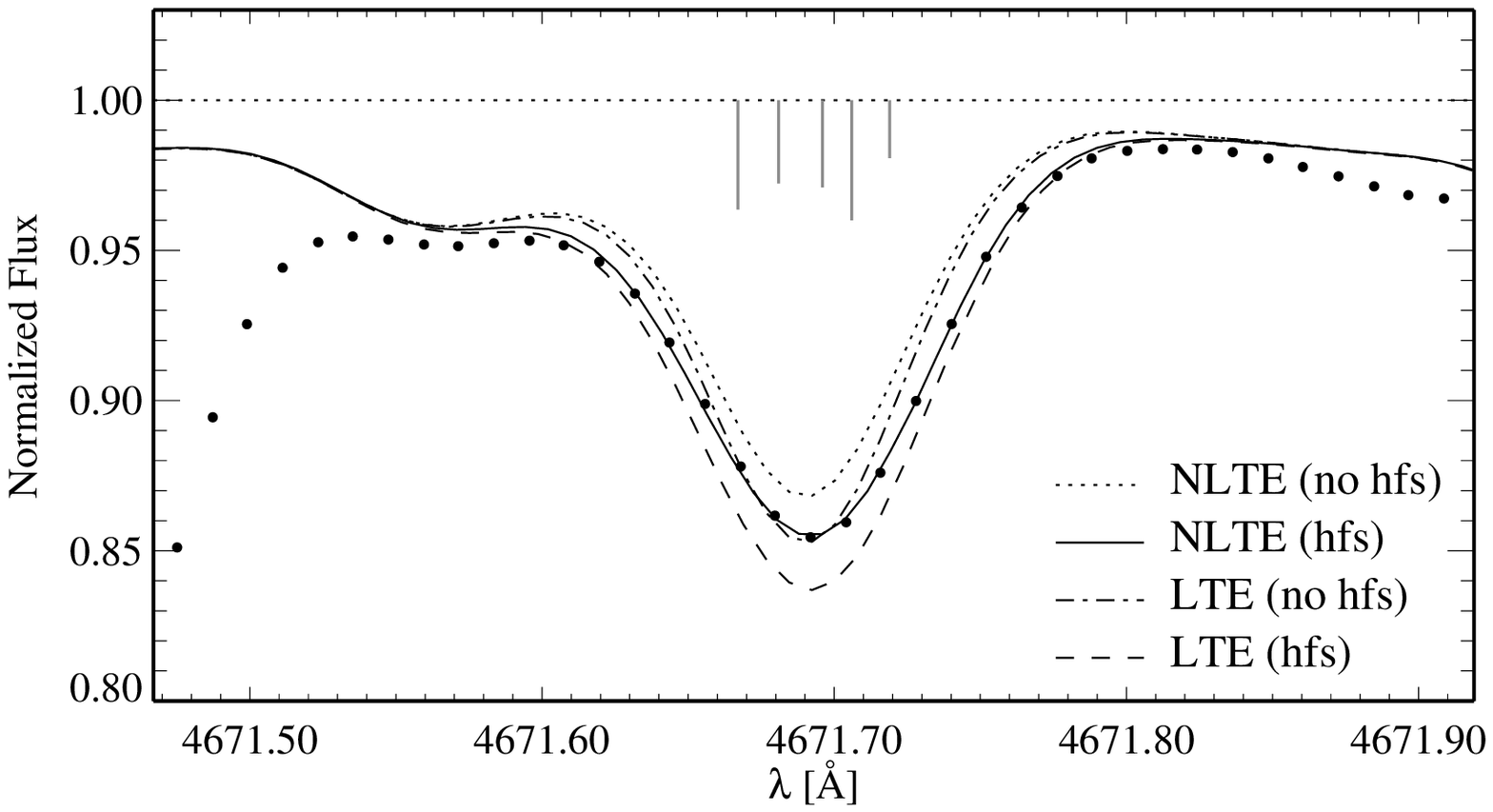}} \hfill
      \resizebox{\columnwidth}{!}{\includegraphics{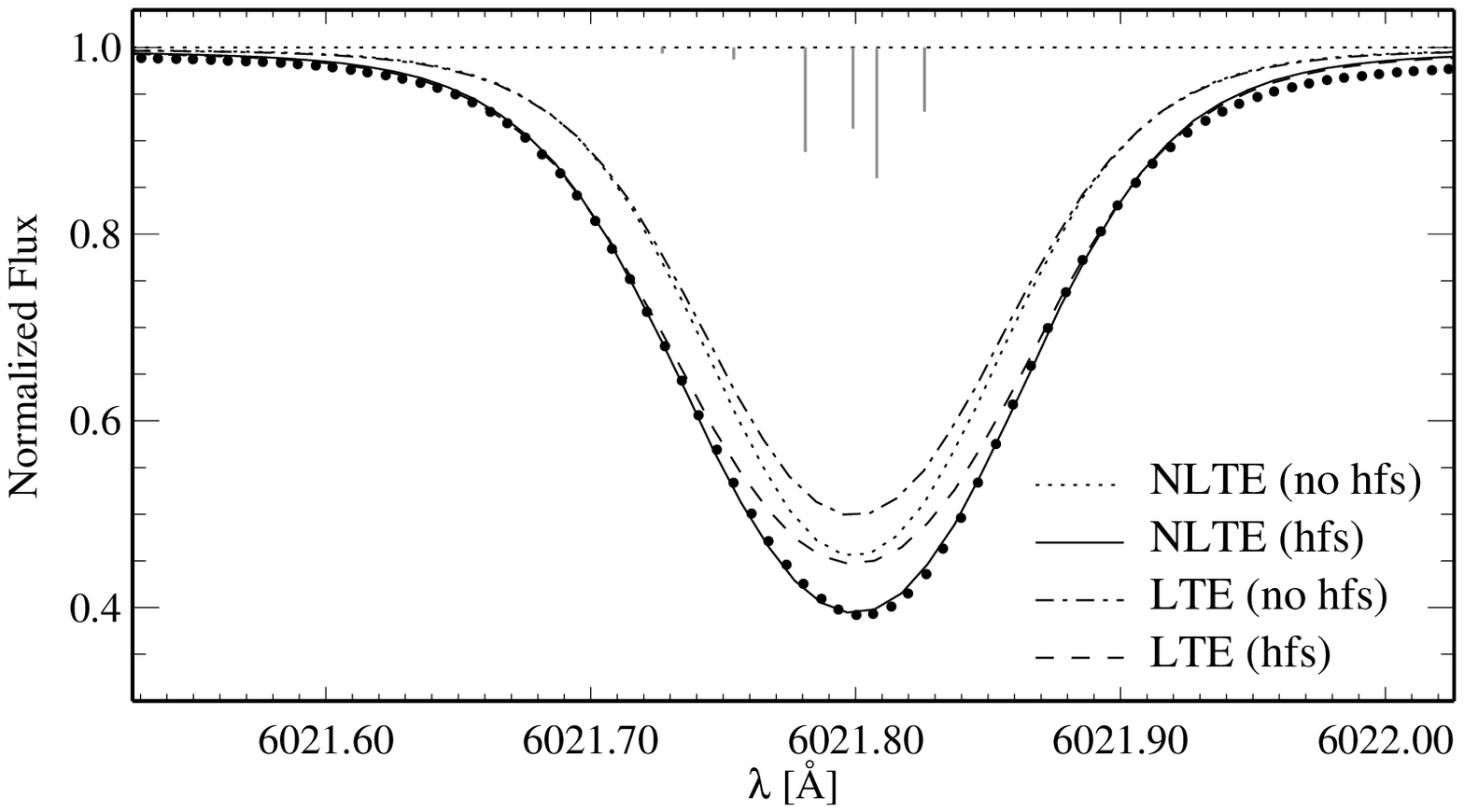}}}
\vspace{-4mm}
\hbox{\resizebox{\columnwidth}{!}{\includegraphics{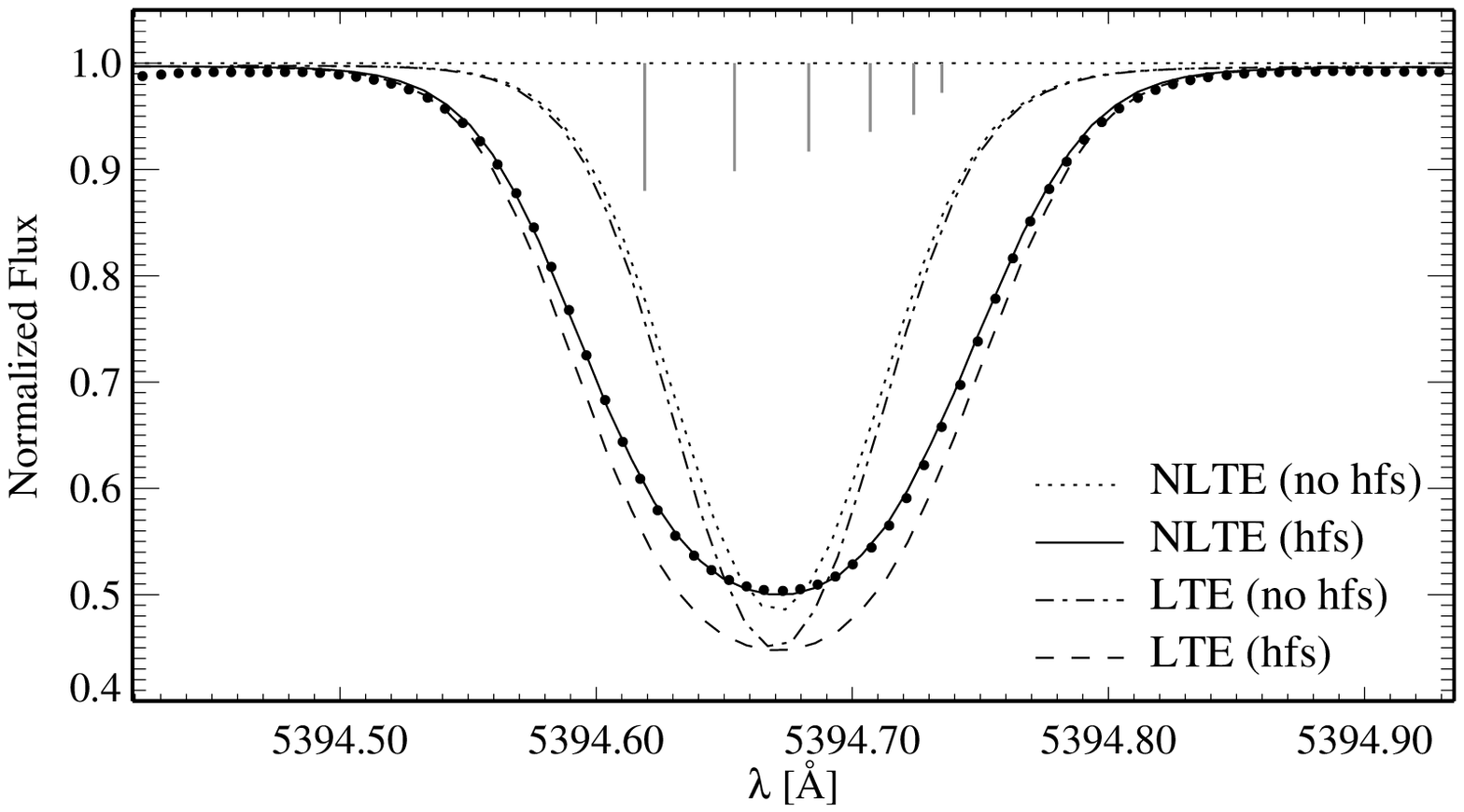}} \hfill
      \resizebox{\columnwidth}{!}{\includegraphics{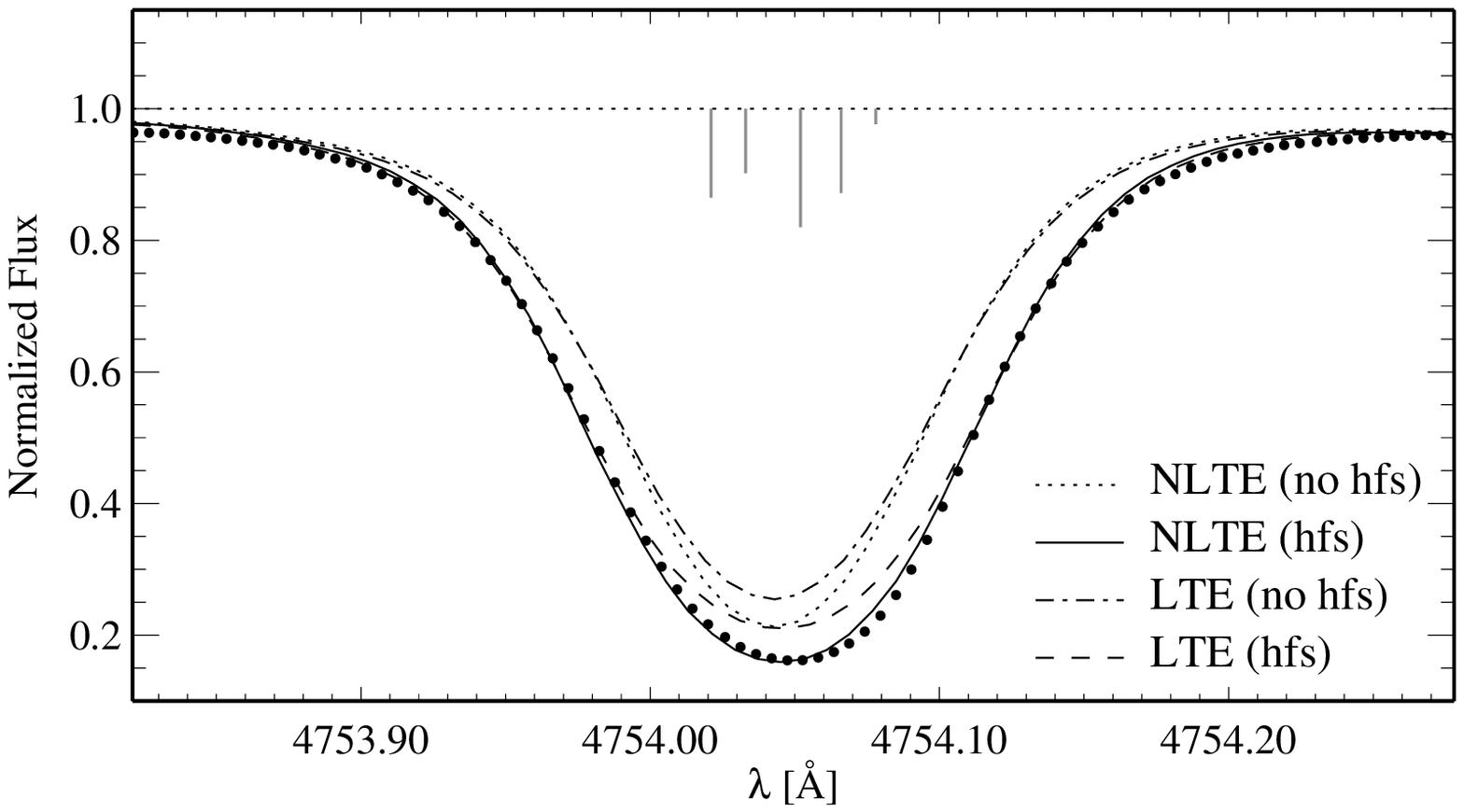}}}
\vspace{-0mm} \caption[]{Selected observed solar \ion{Mn}{i} profiles (filled
circles). Synthetic NLTE and LTE profiles with and without HFS are labeled
correspondingly. NLTE results are based on the reference atomic model.
Wavelength positions and relative line strengths of the HFS components are
indicated. The line abundances and van der Waals' damping constants $\log
C_{\rm 6}$ are $-0.02$,$-31.41$ (5516 $\AA$); $-0.32$,$-30.78$ (4783 $\AA$);
$-0.13$,$-30.73$ (4671 $\AA$); $-0.27$,$-30.64$ (6021 $\AA$); $-0.11$,$-31.40$
(5394 $\AA$); and $-0.16$,$-30.80$ (4754 $\AA$), where the line abundance
parameter refers to the abundance difference with respect to $\logeMn\ = 5.47$
dex.} \label{ltenlte} 
\end{figure*}

The solar spectrum is calculated using the MAFAGS-ODF solar model atmosphere
with $T_{\rm eff} = 5780$ K, an initial manganese abundance\footnote{Note that
the atmospheric model of the Sun does \emph{not} depend on the manganese
abundance.} of $\logeMn\ = 5.47$ dex, and a constant microturbulence velocity
$\xi_{\rm t} = 0.9$ \kms. The profiles are broadened by a rotation velocity
$V_{\rm rot} = 1.8$ \kms, and by a macroturbulence velocity $V_{\rm mac}= 2.5
\ldots 4$ \kms. The observed spectrum is taken from the Kitt Peak Solar Flux
Atlas (Kurucz et al. \cite{Kurucz84}). For the abundance analysis we
originally selected 39 \ion{Mn}{i} lines that satisfy the following conditions:
they are relatively free from blends, and laboratory data for hyperfine
splitting and oscillator strengths are available. Relative intensities of the
HFS components are calculated according to the tables of White \& Eliason
(\cite{WE33}). Magnetic dipole splitting constants, $A(J)$, and electric
quadrupole splitting constants, $B(J)$, of the corresponding levels are
given in the online table. Van der Waals' damping constants $\log C_{\rm 6}$ are
computed according to Anstee $\&$ O'Mara's (\cite{Anstee95}) formalism; a
correction of $\Delta \log C_{\rm 6} = -0.1$ is applied in order to fit the
wings of strong lines. We use oscillator strengths from various sources giving
preference to measured values. All input parameters necessary to perform
spectrum synthesis for the lines of interest are given in Table \ref{lines}.
Line profiles are computed under both LTE and NLTE assumptions; they are fitted
to the observed profiles by means of manganese abundance variations. The
logarithmic abundance differences of the fits with respect to the initial
manganese abundance,  $\Delta \log \varepsilon$, are reproduced in Table
\ref{lines} (Col. 11). In this paper we refer to this parameter as
\emph{abundance correction}. The difference in abundances required to fit LTE
and NLTE profiles is referred to as the \emph{NLTE correction} ($\Delta X =
\log\varepsilon^{\rm NLTE} - \log\varepsilon^{\rm LTE}$); it is given in Col. 12
of Table \ref{lines} for each line.
\subsection{Line profiles}
The computed profiles for selected lines are given in Fig. \ref{ltenlte}
together with the solar spectrum. For comparison we show profiles under LTE and
NLTE conditions, with and without HFS. The examples on the left and on the right
panel are representatives of two groups of lines that we were able to
distinguish in the analysis.

The first group contains those lines, mainly weak with $W_{\lambda} < 80\ \mA$
that are formed deep in the photosphere, in the layers with a strong gradient of
the local temperature $\Te$. The formation of lines under NLTE takes place at
higher $\Te$ when compared to the case of LTE. As $b_i < b_j$ over the entire
line formation depth, the line source function is larger than its LTE value,
$S_{ij} > B_{\nu}(\Te)$. Consequently, NLTE corrections $\Delta X$ will be
positive due to decreased absorption over the entire line profile. In this case,
both LTE and NLTE profiles will fit the observed lines, provided a certain
abundance correction is performed. Of course, abundances obtained from the LTE
profile fitting will be erroneous.

Stronger lines are formed within very large depth ranges. At the depths of line
core formation the upper levels are depopulated more efficiently than the lower
ones, $b_i > b_j$. The source function drops below its LTE value, $S_{ij}~ <
~B_{\nu}(\Te)$, which is driven by photon escape in the line wings. In the lower
atmosphere, behaviour of the source function is inverted, i.e., $b_i < b_j$.
Accordingly, for the strongest lines (e.g., $\lambda\lambda\ 4055, 4783, 6021\ 
\AA$) we observe an amplified absorption in the core and a decreased absorption
in the wings relative to LTE. It is important that profiles computed under the
LTE approach can not fit the observed lines due to their different profile
shapes.

It is noteworthy that we succeeded to fit all lines in the list with a single
value of microturbulence velocity $\xi_{\rm t}$. Although it is well-known that
this parameter approximates a depth-dependent velocity field in the solar
atmosphere, we decided to preserve a constant $\xi_{\rm t} = 0.9$ \kms.
Certainly, this assumption does not always give the best result: lines of
intermediate strength require different values of the microturbulent velocity to
fit the core and the wings simultaneously. We have not tried to adjust this
value. Our analysis of the solar lines is mainly aimed at derivation of atomic
and atmospheric data to be used in metal-poor stars, where only a single
depth-independent value of $\xi_{\rm t}$ can be recovered. Other asymmetries in
the observed lines, such as red wing deficit or core skewness (as seen on all
lines in the right panel), are a reflection of atmospheric inhomogeneities,
which we can not take into account with our static plane-parallel models.
\subsection{Solar abundance of Mn and oscillator strengths}
\begin{figure}
\resizebox{\columnwidth}{!}{\includegraphics{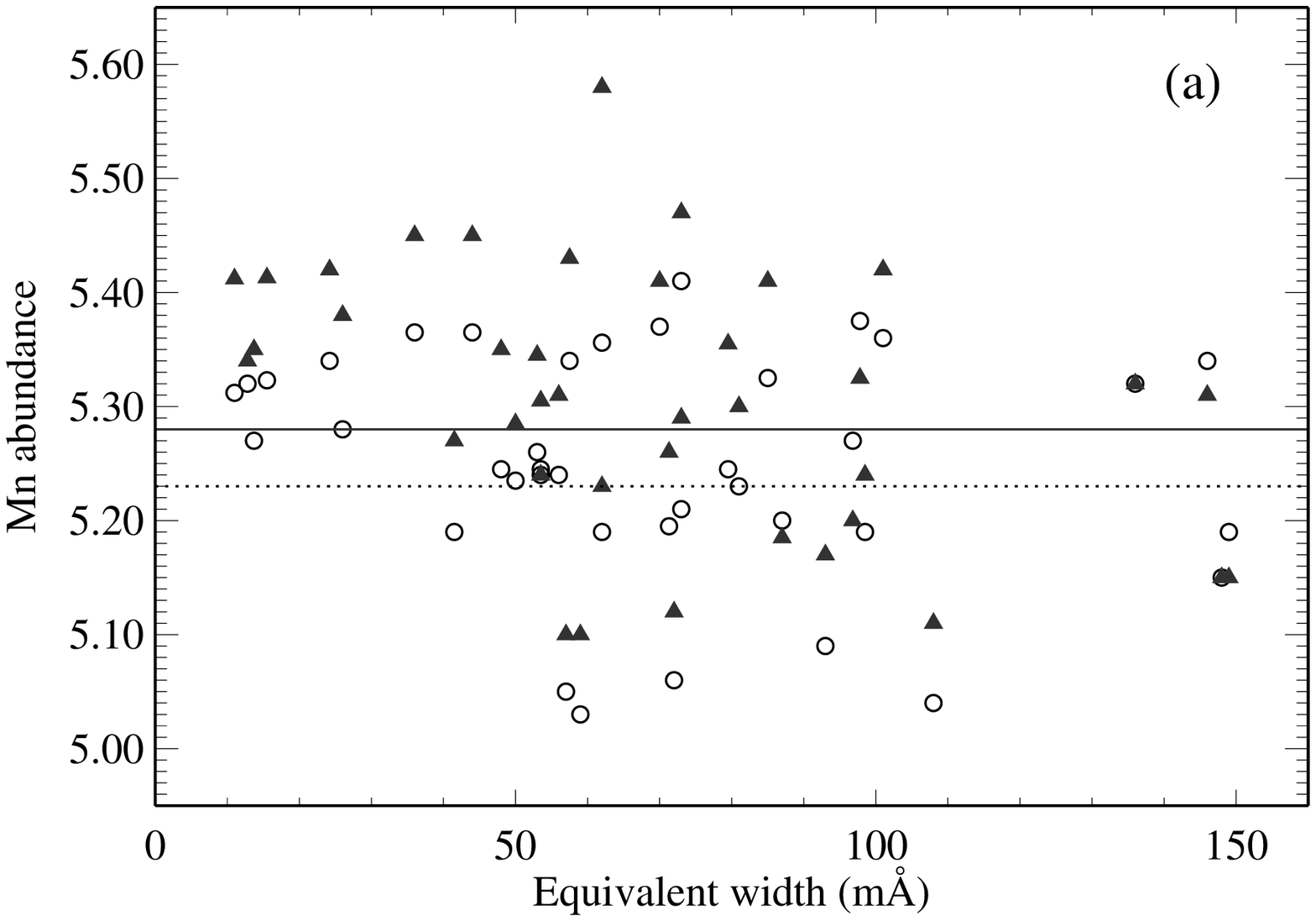}}
\resizebox{\columnwidth}{!}{\includegraphics{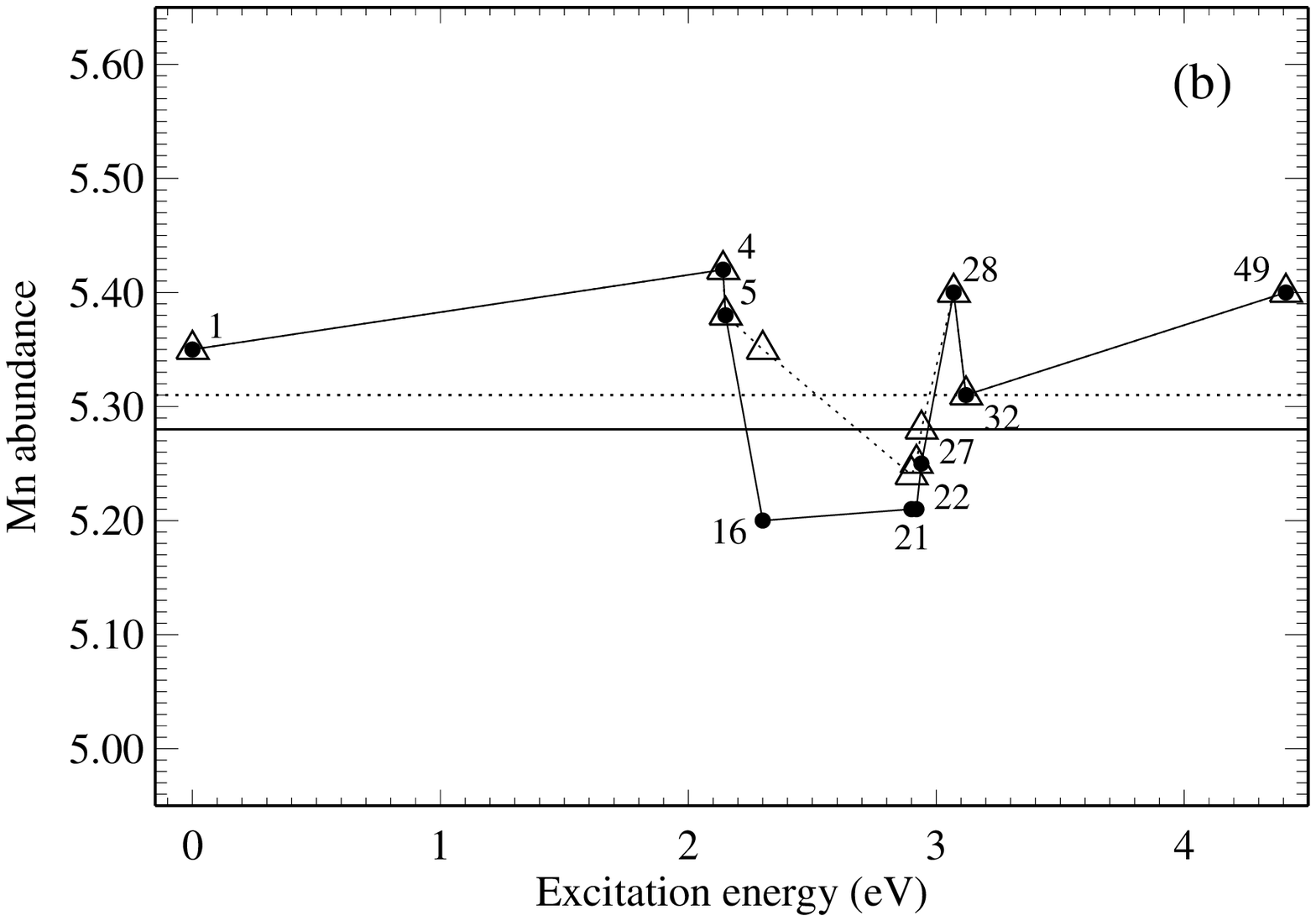}}
\resizebox{\columnwidth}{!}{\includegraphics{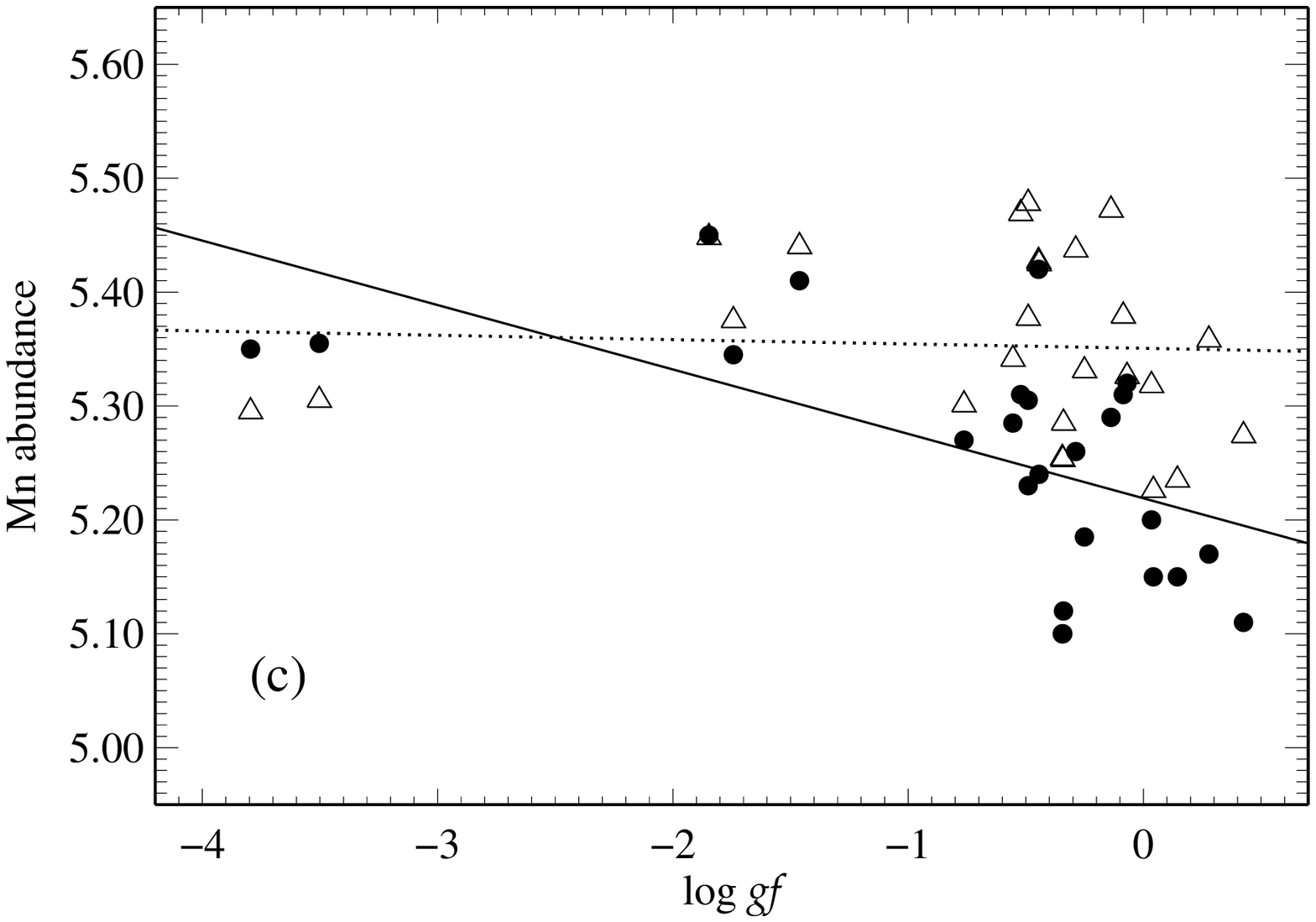}}
\caption[]{Abundances of the 39 \ion{Mn}{i} lines in Table \ref{lines}.
(a): LTE (circles) and NLTE (filled triangles) abundances. The mean LTE
abundance is represented by dots, NLTE by a continuous line. (b): Mean
abundances in \emph{multiplets}. Filled circles refer to the standard NLTE
calculations. Open triangles refer to the NLTE calculations with adjusted
$\log C_6$ values (see text). The solid black horizontal line indicates the
mean multiplet NLTE abundance, $\logemean\ = 5.28$, for all lines without $\log
C_6$ correction, the dotted horizontal line represents a corresponding value of
$\logemean\ = 5.31$ dex after $\log C_6$ adjustment. Multiplet numbers are
indicated. (c): NLTE abundances as a function of oscillator strengths. Filled
circles and a continuous linear regression line denote Oxford oscillator
strengths of Booth et al. (\cite{Booth84a}), open triangles and a dotted line
refer to Kurucz's database.}
\label{abundall}
\end{figure}

The method of spectrum synthesis yields the product of the oscillator strength
for a transition and the abundance of the element, $\loggfe$. This
parameter, as well as NLTE correction $\Delta X$, and abundance correction
$\Delta\log {\varepsilon}$, are given in Cols. 13, 12, and 11 of Table
\ref{lines}, respectively. Figure \ref{abundall}a shows NLTE and LTE abundances
for all lines as a function of their equivalent widths. It is seen that the
absolute abundances determined assuming LTE are \emph{lower} than those derived
with NLTE level populations. We find the weighted\footnote{\emph{All} mean
abundances in this paper are weighted according to the uncertainties in the
$gf$-values for all lines; higher weights are assigned to lines with smaller
errors} mean NLTE and LTE abundances $\logeMnN\ = 5.28 \pm 0.11$ dex and
$\logeMnL\ = 5.23 \pm 0.1$ dex. The standard deviations $\sigma $ are quoted as
errors.

It is important to note that there is no single set of measured oscillator
strengths for all lines in our list. Data from five sources were used, and this
undoubtedly reduces the precision of our results, leading to a large
standard deviation of the mean abundance. As a test, we calculated the
mean $\logeMnN\ $ for \emph{each} set of $gf$-values (Table \ref{ablogf}). The
result is a large spread of abundances from 5.23 to 5.46 dex, which is due to
strongly discriminate oscillator strengths for the same lines measured by
different authors. Indeed, for the majority of lines in our list the mean
difference $\log gf $(Oxford -- Kurucz) = 0.096 dex, $\log gf $(Oxford --
Becker) = 0.09 dex. There is, surprisingly, much better agreement between the
absolute oscillator strengths of Becker et al. (\cite{Becker80}) and Kurucz
(\cite{Kurucz88}): the mean difference (Oxford -- Becker) = 0.047 dex.
We have, nevertheless, serious doubts about the accuracy of all
measurements: the errors, quoted by the authors (0.04 to 0.1 dex), are much
smaller than the actual differences between the $gf$-values from these sources.

Only 14 lines in our list (marked in the Table \ref{lines} with an asterisk)
have consistent measured oscillator strengths. When only those lines are used,
we obtain $\logeMnN\ = 5.30 \pm 0.09$ dex. When two obvious ``outsiders'' with
$\loge\ = 5.15$ dex from multiplet 16 are ignored, the abundance of manganese
increases to $5.36 \pm 0.06$ dex. Despite its large standard deviation, this
value based on only 12 lines is more reliable than the one calculated with
\emph{all} lines from our sample.

\begin{table}
\caption[]{Weighted NLTE manganese abundances based on oscillator strengths from
different sources.}
\label{ablogf}
\begin{tabular}{cclccl}
\hline\noalign{\smallskip}
$N_{\rm lines}$ & Source of & Method & accuracy & $\loge$ & $\sigma$ \\
          & $\log gf\,^a$ &        & $\%$       &  [dex]    & \\
\noalign{\smallskip}\hline\noalign{\smallskip}
17 & 1 &  beam foil        & 25 & 5.35 & 0.07\\
20 & 2 &  beam foil        & 50 & 5.46 & 0.2 \\
19 & 3 &  laser excitation &    & 5.30 & 0.08\\
25 & 4 &  total absorption & 10 & 5.23 & 0.10\\
39 & 5 &  calculations     &    & 5.37 & 0.08\\
39 & all &                 &    & 5.28 & 0.11\\
\noalign{\smallskip}\hline\noalign{\smallskip}
\end{tabular}
{\footnotesize
\ $^a$\ References:\ \  (1) Greenlee \& Whaling \cite{Green79}; (2) Woodgate
\cite{Wood66}; \\
   (3) Becker et al. \cite{Becker80}; (4) Booth et al. \cite{Booth84a}; (5)
Kurucz \cite{Kurucz88}}
\end{table}
\begin{table}
\caption{NLTE abundances $\loge$ of \ion{Mn}{i} \emph{multiplets}. Values
obtained with $\log gf$-values from different sources are marked with an
asterisk; $\Delta \loge $ values are given relative to the \emph{weighted mean}
NLTE abundance $\logeMnN\ = 5.28$ dex.}
\label{groups}
\begin{tabular}{rcllllrc}
\noalign{\smallskip}\hline\noalign{\smallskip}
Mult. & $N_{lines}$ & $\Elow$ & $\log C_{\rm 6}$ & $\loge$ &
  $\sigma$ & $\Delta \loge $\\
 &  & [eV] &  & [dex] &  &  & \\
\noalign{\smallskip}\hline\noalign{\smallskip}
1  &  2  &  0.0   &  --31.4   & 5.35  & 0     &  0.07 \\
4  &  5  &  2.14  &  --31.41  & 5.42* & 0.04  &  0.14 \\
5  &  3  &  2.15  &  --31.0   & 5.38  & 0.06  &  0.1~~  \\
16 &  3  &  2.3   &  --30.8   & 5.20  & 0.09  &--0.08 \\
20 &  1  &  2.91  &  --30.83  & 5.35  &       &  0.07 \\
21 &  7  &  2.9   &  --30.74  & 5.21* & 0.09  &--0.07 \\
22 &  7  &  2.92  &  --30.67  & 5.21  & 0.10  &--0.07 \\
23 &  1  &  2.94  &  --30.4   & 5.58  &       &  0.2~~  \\
27 &  3  &  3.06  &  --30.64  & 5.25* & 0.08  &--0.02 \\
28 &  2  &  3.07  &  --30.7   & 5.40* & 0.13  &  0.12 \\
32 &  2  &  3.12  &  --30.73  & 5.31* & 0.11  &  0.03 \\
49 &  3  &  4.42  &  --31.6   & 5.40  & 0.02  &  0.12 \\
\noalign{\smallskip}\hline\noalign{\smallskip}
\end{tabular}
\end{table}

As can be seen from Table \ref{lines}, there is a large line-to-line spread of
abundances. To investigate this we calculated mean abundances $ \loge$ and
standard deviations for each \emph{multiplet} (Table \ref{groups}).
The abundances obtained with oscillator strengths from different sources are
marked with an asterisk in the abundance column; these values are also
weighted according to the errors in oscillator strengths. A plot of $\loge$ is
shown in Fig. \ref{abundall}b. There is a large difference between $\loge$ for
multiplets with $\Elow = 2 \ldots 3 $ eV and all other lines. It is interesting 
that a similar irregularity was noted by Blackwell et al. (\cite{Black82}) and
Simmons \&\ Blackwell (\cite{Simmons82}) from the analysis of \ion{Fe}{i} lines
in the solar spectrum. Lines with excitation energies 2.18 \ldots 2.2 eV gave
lower abundances than lines with smaller or larger $\Elow$. Inaccurate damping
parameters (Uns\"old formula) or a vague dependence on the multiplicity of a
term were suggested as the cause.

The only common feature between our analysis and the iron results is the use of
oscillator strengths measured by the Oxford group. These laboratory data cover
61\%\ of all lines in our list; in particular, abundances for multiplets 16 and
22 are calculated solely with $\log gf $ values from Booth et al.
(\cite{Booth84a}). Standard deviations for these multiplet abundances are no
larger than in other multiplets, although Fig. \ref{abundall}c shows a weak
correlation of abundances with the oscillator strengths, where the correlation
coefficient is equal to --0.7. On this plot, abundances calculated using Booth
et al.'s (\cite{Booth84a}) data are compared with those calculated with
Kurucz's data. The linear regression curves are shown for both sets. However,
the observed behaviour could be a random variation misinterpreted for a trend,
or it may be the consequence of incorrectly chosen damping parameters.

\begin{figure}
\resizebox{\columnwidth}{!}{\includegraphics{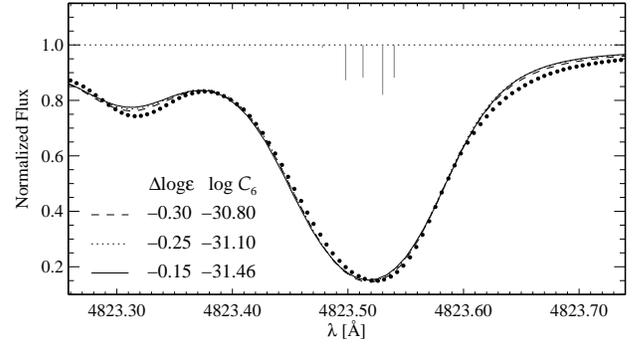}}
\caption[]{NLTE line profiles of the line 4823 \AA. Different combinations of
damping parameters and resulting abundance deviations from
$\logeMn\ = 5.47$ dex are shown.}
\label{c6profile}
\end{figure}

In order to test the influence of the Van der Waals damping we carried out
spectrum synthesis calculations adjusting $C_{\rm 6}$ for each multiplet. The
correction of this parameter was found from the requirement of obtaining equal
abundances for weak and strong lines of a common multiplet; $\Delta \log C_{\rm
6}$ for all investigated multiplets averaged to --0.7 dex relative to the Anstee
\&\ O'Mara data, which is close to values predicted by the Uns\"old formula.
However, \emph{only} the lines of intermediate multiplet numbers demonstrated a
weak sensitivity to the Van der Waals damping (Fig. \ref{abundall}b): $\Delta
\loge$(16)$ = +0.14$ dex, $\Delta \loge$(21)$ = +0.03$ dex, $\Delta \loge$(22)$
= +0.04$ dex, $\Delta \loge$(27)$ = +0.03$ dex. This is insufficient to explain
\emph{the discrepancy of 0.2 dex} between the abundances from lines of these
multiplets and the others. Besides, it is not always possible to discern the
influence of $C_{\rm 6}$ and $\loge$ on the line profiles. This is demonstrated 
in Fig. \ref{c6profile}, where three combinations of both parameters, i.e.,
smaller abundance and larger damping, or larger abundance and smaller damping,
lead to equally good fits. The difference between the profiles is visible only
in the blend on the blue line wing. In our final estimate of the solar Mn
abundance we decided to keep the Anstee \&\ O'Mara damping parameters, even
though they require a significantly smaller abundance for particular lines.

The discrepancy between the abundances (both NLTE and LTE) from different
multiplets is hard to explain. We do not believe that errors in damping could be
 as large as this, neither with confidence can we attribute this to errors in
the oscillator strengths. It is interesting that a similar result was obtained
for \ion{Mn}{i} lines by the other authors. Lines that give a strong
underabundance in our analysis also have smaller $\loge$ relative to the
average value in the work of Becker et al. (\cite{Becker80}). Booth et al.
(\cite{Booth84b}) also distinguished two groups of lines based on the lower
excitation potential: the 0 eV lines, which show considerably higher abundance,
and those between 2 -- 3 eV with lower abundances, but are sensitive to damping
enhancement. To resolve this inconsistency we need new quality measurements of
oscillator strengths and, perhaps, a depth-dependent microturbulence or a 3D
hydrodynamical model.
\begin{figure}
\resizebox{\columnwidth}{!}{\includegraphics{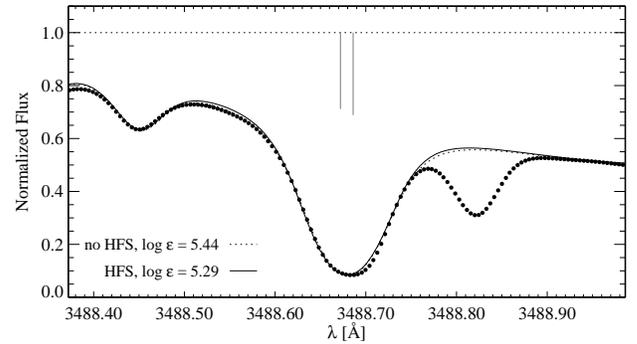}} \caption[]{LTE
line profiles of the \ion{Mn}{ii} line at 3488 \AA. The strong line wings
centered near 3489.05 \AA\ are most likely due to an autoionization line, 
represented as an \ion{Fe}{i} transition at 3.5 eV with an extremely large
radiative damping constant.} \label{mnii3488}
\end{figure}

Unfortunately, we can not solve the main problem: there is a strong
inconsistency between the solar photospheric abundance of Mn and a meteoritic
value of $5.50 \pm 0.03$ dex (Lodders \cite{Lodders03}). Quite on the contrary
to the expectations, every subsequent analysis with more precise atomic data and
model assumptions gives progressively \emph{decreasing} $\logeMn\ $. Generally,
high values that are consistent with the meteoritic abundance (e.g., Greenlee
$\&$ Whaling \cite{Green79}; Becker et al. \cite{Becker80}) result from
erroneous oscillator strengths, neglect of HFS, an LTE approach (our work is the
\emph{first} NLTE analysis), and/or unaccounted blends. For instance, Greenlee
$\&$ Whaling (\cite{Green79}) used only six lines with $W_\lambda < 14\ \mA $,
assuming that there are no unrecognized blends in the solar spectrum and
neglecting HFS. Yet, we have found that four of these six lines contain blends, 
we therefore ignored them for our analysis. The most recent and commonly used
LTE abundance of Mn ($5.39 \pm 0.03 $ dex) is that of Booth et al.
(\cite{Booth84b}). It is also rather low, hence the authors suggested that
NLTE effects could be the cause of methodical errors. Indeed, NLTE effects of
\ion{Mn}{i} in the Sun \emph{are} significant, but they only increase the
existing discrepancy. Even if only the weak lines with $W_\lambda < 50\ \mA$ are
used, which are relatively insensitive to the \textit{keystones} of the
abundance analysis (damping, model structure, HFS, microturbulence), we obtain a
weighted NLTE abundance of 5.37 $\pm$ 0.06 dex. The incorrect abundance of the
reference element Si may be the cause. Some authors compare the abundance of
manganese with that of iron. The recent NLTE estimate is $\logeFe\ = 7.48
\ldots 7.51$ dex (Gehren et al. \cite{Gehren01b}). If we adopt this value, a
ratio Fe/Mn = 148 is derived; it is significantly larger than Fe/Mn = 89 in CI
chondrites (Cameron \cite{Cameron73}). Perhaps, as suggested by Booth et al.
(\cite{Booth84b}), the chemical composition of CI meteorites does not fully
represent the composition of the proto-Sun.

To investigate the ionization equilibrium of manganese, we have tried to
determine the abundance from \ion{Mn}{ii} lines. Martinson et al.
(\cite{Martinson77}) claim that \ion{Mn}{ii} lines can be used to obtain the
same photospheric solar abundance as found from \ion{Mn}{i} lines. From
equivalent width measurements of 9 lines of ionized manganese these authors have
calculated $\logeMn\ = 5.4 \pm 0.2$ dex. We have checked the solar spectrum of
\ion{Mn}{ii} and found that no lines are suitable for any of the abundance
determination methods. All lines strong enough to be discerned in the solar
spectrum are located in the near-UV, and are heavily blended, which complicates
the continuum placement. We have, nevertheless, synthesized these lines under
the assumption of LTE: all lines of interest have saturated cores, i.e. the
variation of the abundance within $\pm$ 0.2 dex affects only the strength of the
wings. Thus, due to the presence of strong blends in the line wings, no
conclusion can be made about the true abundance. For the line at $\lambda$ 3488
$\AA$ the effect of HFS was investigated (see Fig. \ref{mnii3488}). Data for the
levels involved in this transition were taken from Holt et al. (\cite{Holt99}).
Hyperfine structure broadens the profile and leads to a better fit of the inner
wings provided the microturbulence is reduced to $\xi_{\rm t} = 0.8$ \kms. Its
effect is very significant: a line profile calculated with no HFS and $\logeMnL\
= 5.44$ dex is equal to a profile with 2 HFS components and $\logeMnL\ = 5.29$
dex. Hence, abundances for the \ion{Mn}{ii} lines calculated without accounting
for HFS can easily be overestimated by 0.2 dex. Our results bring into
perspective the conclusions of Martinson et al. (\cite{Martinson77}) : the
neglect of HFS in their calculations is, most likely, responsible for an
overestimated abundance of manganese. At the same time, the comparison of
abundances derived from two ionization stages of manganese turns out to be an
unreliable technique. It should be performed with caution.
\subsection{Uncertain parameters}
\subsubsection{Photoionization and collisions with \ion{H}{i} atoms}
We performed NLTE calculations with various scaling factors for photoionization
and collision cross-sections. The effect on the departure coefficients is
described in detail in Sect. 3.2. Here we show how changing these parameters
affects the line profiles and the resulting abundances.

\begin{figure}
\resizebox{\columnwidth}{!}{\includegraphics{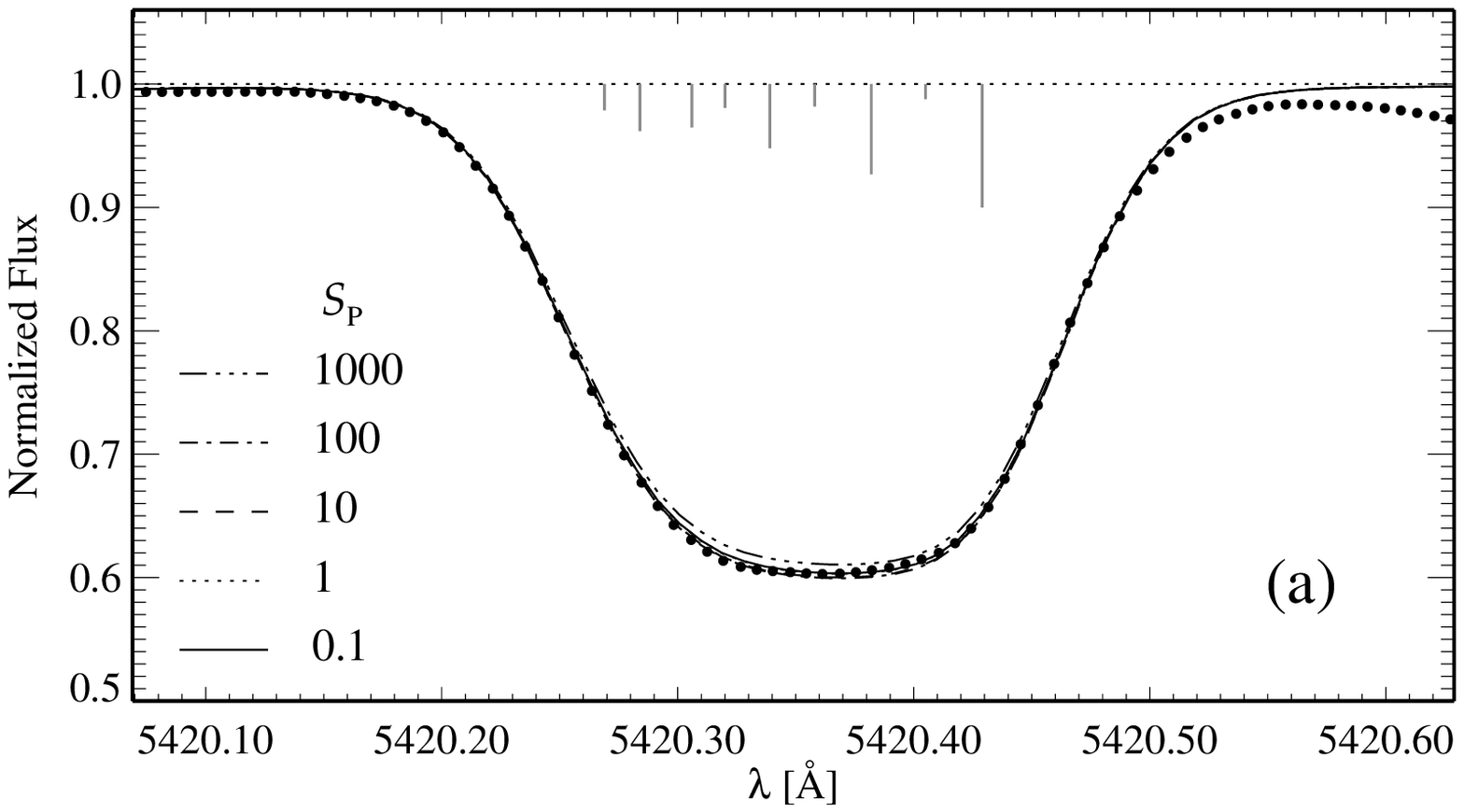}}
\resizebox{\columnwidth}{!}{\includegraphics{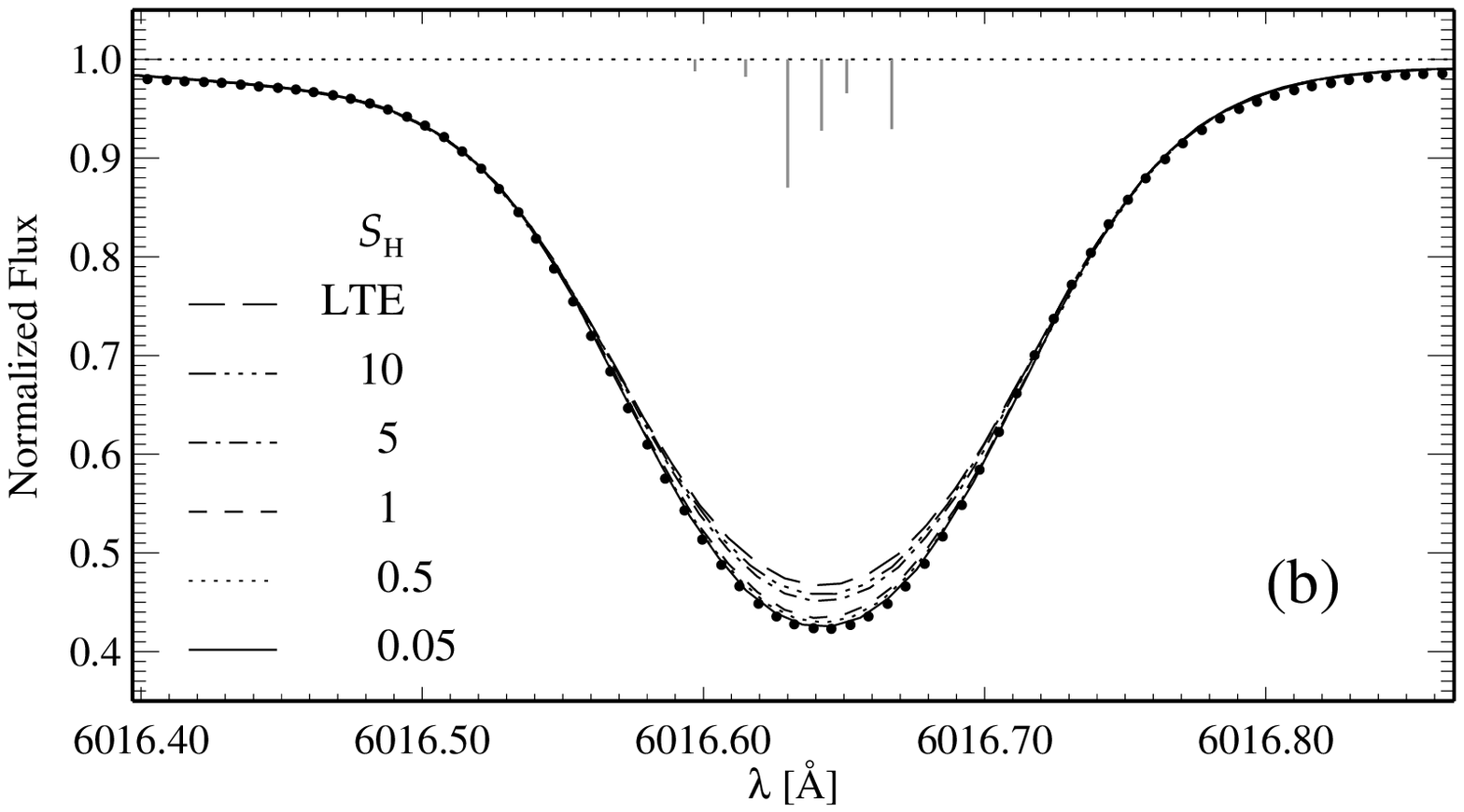}}
\resizebox{\columnwidth}{!}{\includegraphics{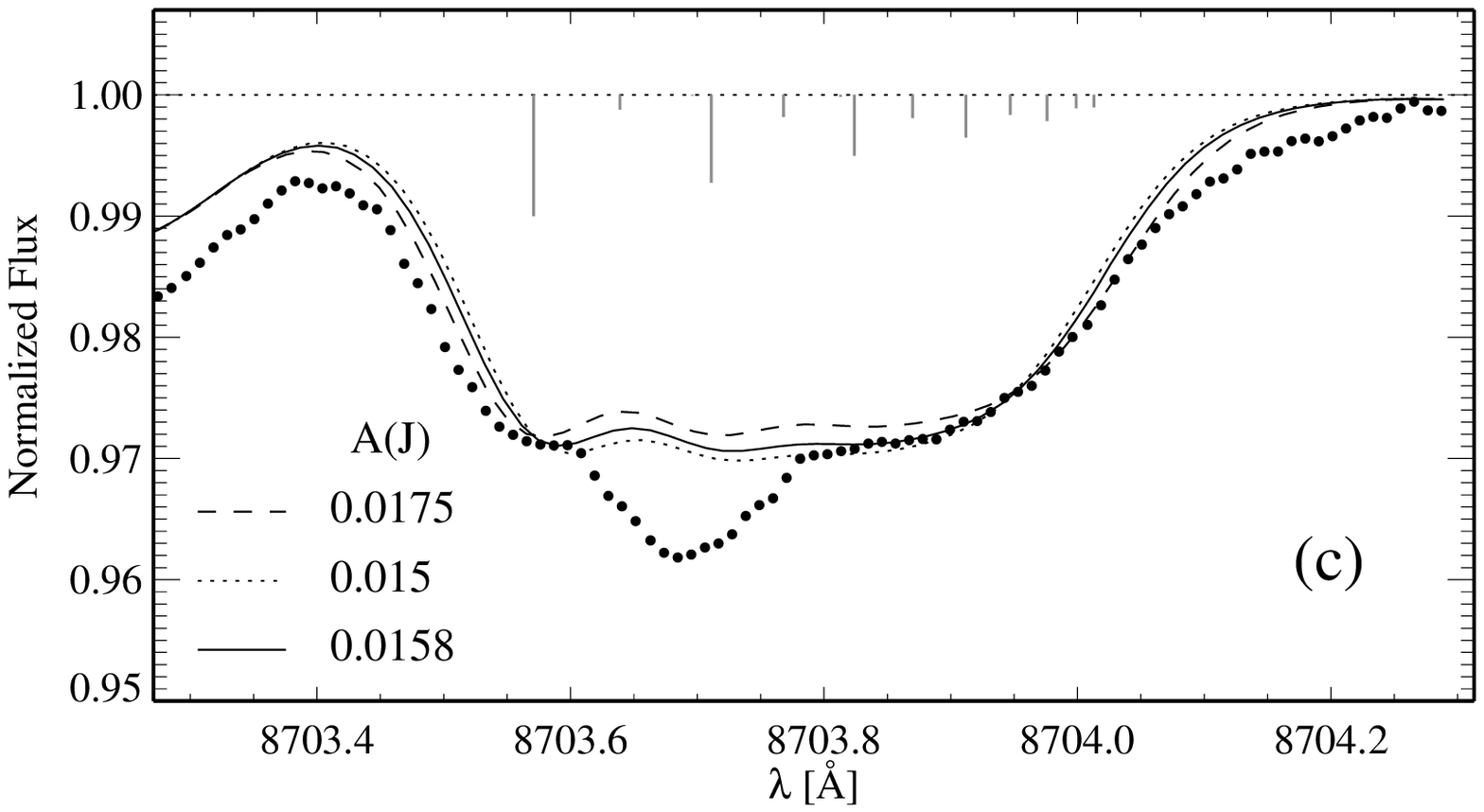}}
\caption[]{Synthetic NLTE profiles for the \ion{Mn}{i} lines. (a):
Variation of $\lambda 5420$ \AA\  with photoionization cross-sections multiplied
by different scaling factors $\SP$. (b): Response of $\lambda 6016$ \AA\ to
various scaling factors for \ion{H}{i} collisions. (c): The \ion{Mn}{i} line
8703 \AA\ displaying the dependence on the HFS constant $A(J)$. The line suffers
from an unknown blend.} \label{parvar}
\end{figure}

The behaviour of profiles for different scaling factors $\SP$ to hydrogenic
photoionization cross-sections is shown in Fig. \ref{parvar}a. The difference in
the profiles is only marginal: maximum abundance corrections amount to -0.04 dex
with $\SP$ increasing from 0 to 1000. This effect was confirmed with twelve
lines. We conclude that the photoionization cross-sections are not
crucial in \emph{abundance calculations} as long as very large $\SP$
enhancements (by a few orders of magnitude) to hydrogenic cross-sections are
avoided.

A similar test was performed for collisions with \ion{H}{i} atoms, the parameter
usually considered the most uncertain in NLTE calculations. Scaling factors $\SH
= 0.05, 0.5, 1, 5$ and $10$ were applied to b-b and b-f cross-sections computed
from Drawin's formula. The NLTE profiles of the line at 6016 $\AA$ are presented
for five different assumptions in Fig. \ref{parvar}b. As expected, with
increasing rates of collisional excitation and ionization by \ion{H}{i} atoms,
the NLTE profile approaches LTE. However, the actual correction to the abundance
is rather small: for ten lines investigated it does not exceed 0.06 dex for $\SH
= 5$. Neglect of hydrogen collisions (not shown here) requires a reduction of
the abundance by 0.01 dex. According to the results of Belyaev \&\ Barklem
(\cite{Belyaev03}), Drawin's formula ($\SH = 1$) strongly overestimates the rate
of transitions due to collisions with \ion{H}{i}. Therefore, this scaling factor
is already too large, and we conclude that the magnitude of uncertainty
concerned with the choice of $\SH$ is not much larger than 0.05 dex.
\subsubsection{Hyperfine splitting data}
The sensitivity of lines to \emph{uncertainties} in the hyperfine splitting data
was investigated (Fig. \ref{parvar}c). It is seen that differences in the 8703
$\AA$ line profiles computed for $A(J) = 0.015$ cm$^{-1}$, $A(J) = 0.0158$
cm$^{-1}$, and $A(J) = 0.0175$ cm$^{-1}$ are rather small. These differences
correspond to an abundance correction of 0.02 dex. A notable change in the
profile is found only when $A(J)$ is modified by more than 4 mK ($1 \rm{mK} =
0.001$ cm$^{-1}$). However, the precision of $A(J)$ measurements that we use is
very high; the average error is 0.5 mK. We have tested other lines, including
those with saturated cores, varying the $A(J)$ value for the corresponding
levels by $\sim$ 2 mK. In this case, the relative shift of the components leads
to $\Delta \log\varepsilon \sim 0.01 - 0.02$ dex. We conclude that due to the
much stronger influence of the other parameters ($\log gf $, $C_{\rm 6}$),
uncertainties in HFS data do not affect the resulting abundances. However,
correcting the $A(J)$ value may in a few cases help to obtain a better fit to
the observed spectral lines.

The influence of HFS on line formation depths is interesting. The effect of
including HFS on the position and width of formation layer for the weak
resonance line 5394 $\AA$ was investigated by Vitas (\cite{Vitas05}). The main
result was that the line was formed higher in the photosphere when HFS was not
taken into account. We have tested this observation on a sample of lines with
different $W_\lambda$ and $gf$ values; their depths of formation computed with
and without HFS are given in Table \ref{depform}. These are average depths of
NLTE line formation calculated with the contribution function to the total
emergent radiation according to Achmad et al. (\cite{Achmad91}).
\begin{table}
\caption{Average depths of \ion{Mn}{i} NLTE line formation for a line with
continuum according to Achmad et al. (\cite{Achmad91}).}
\label{depform}
\begin{tabular}{cc|ccc|c}
\noalign{\smallskip}\hline\noalign{\smallskip}
 & & \multicolumn{3}{c|}{\raisebox{0.5ex}{$\log \tau_{\rm 5000}$}} & \\
$\lambda$ [\AA] & $W_\lambda$, [$\mA$] & HFS & no HFS & wing & $\Delta T$, K\\
\noalign{\smallskip}\hline\noalign{\smallskip}
 5394.67 & 78 & -0.7  & -1.4  & -0.18 & 420 \\
 5432.51 & 48 & -0.57 & -0.86 & -0.19 & 259 \\
 4055.54 & 136& -2.43 & -2.68 &  0.05 & 57 \\
 4058.93 & 101& -1.88 & -2.14 &  0.05 & 63 \\
 4070.28 & 70 & -0.88 & -1.10 &  0.05 & 135 \\
 6013.50 & 87 & -1.03 & -1.41 & -0.26 & 190 \\
 6016.64 & 98 & -1.18 & -1.51 & -0.26 & 137 \\
 6021.80 & 97 & -1.68 & -1.84 & -0.26 & 46 \\
\noalign{\smallskip}\hline\noalign{\smallskip}
\end{tabular}
\end{table}
Both for strong and weak lines the depth of core formation decreased when
hyperfine splitting was neglected in line synthesis. The corresponding shift in
local temperatures $\Delta T$ is given in the last column. We note that the
formation of weaker lines takes place at significantly higher $\Te$. For
instance, the resonance line at 5394.67 $\AA$ with 6 HFS components is formed in
the layer with $\Te = 5340$ K, rather than in the layer with $\Te = 4920$ K
(depth of formation for the same line with a single component). This result
shows that weak lines are to a greater degree affected by \emph{inclusion
of HFS in line synthesis}, compared to stronger lines.
\section{Conclusions}
The NLTE formation of \ion{Mn}{i} lines in the solar atmosphere is
rather complex. Our results indicate that manganese belongs to a class of mixed
domination atoms, where several channels compete in establishing non-equilibrium
level populations. However, \emph{in our model} neither of the processes can be
regarded as a dominating type: in particular, the absence of photoionizations
does \emph{not} restore an LTE distribution of populations. We suggest that the
photoionization inefficiency is the simple result of too low cross-section
approximations. Whether with or without quantum defect, the hydrogenic
approximations deliver values roughly factors of 100 to 1000 lower than seen in
comparable atoms for which better calculations are available (Fe, Si).

The depopulation of the \ion{Mn}{i} ground state and low excitation
levels in the absence of photoionization can be understood as follows: radiative
transitions with strong optical pumping effects cause an overpopulation of the
high-excitation levels with respect to the low-excitation levels. Due to the
strong collisional coupling of the high-excitation levels with the \ion{Mn
}{ii} ground state, which is in LTE, the departure coefficients of these levels
will be close to unity; consequently the departure coefficients of the
low-excitation levels will drop  significantly below unity. In the absence of
the high-excitation levels, \ion{Mn}{i} behaves as a photoionization-type ion.
Disregarding inelastic collisions with electrons results in a weaker coupling of
the levels to each other and a stronger deviation of line source functions from
$B_{\rm \nu}(\Te)$.

All levels that are involved in transitions of interest are underpopulated
relative to LTE. Hence the NLTE assumption generally leads to an increase of the
\ion{Mn}{i} depth of line formation that forces a weakening of the lines.
However, for every particular line the ratio of the upper and lower level
population, determined by an interplay of the above-defined NLTE processes,
modulates the behaviour of the wings and cores, and hence the value and sign
of NLTE corrections. For the weak and intermediate strength lines ($W_{\lambda}
< 80\ \mA$), NLTE and LTE assumptions give a similarly good fit to the
observations, provided a certain abundance correction is performed. The NLTE
corrections for these lines are positive and amount to 0.1 dex. Formation of the
stronger lines ($W_{\lambda} > 80\ \mA$) is characterized by an amplified
absorption in the core and a decreased absorption in the wings, compared to LTE.
NLTE corrections for strong lines are scattered around zero or negative.

The LTE analysis of \ion{Mn}{i} has two pitfalls: the strong lines can not be
reproduced due to different profile shapes, and abundances of the weak lines
under LTE are overestimated. For this reason we do not recommend the commonly
used solar LTE abundance $\logeMn\ = 5.39 \pm 0.03$ dex, obtained by Booth et
al. (\cite{Booth84b}). Instead, an NLTE abundance of 5.36 $\pm$ 0.1 dex,
determined from 12 lines of $W_\lambda = 12 - 140\ \mA$ with relatively reliable
oscillator strengths in the current analysis, is suggested. We consider
abundances from other lines in our list as unreliable due to (1) high
sensitivity of particular lines to $C_{\rm 6}$ (for multiplet 16 $\Delta \loge\
= +0.14 $ dex for $\Delta \log C_{\rm 6} = -0.9$) and (2) inconsistent
oscillator strengths between different sources. This result increases the
discordance with the abundance of Mn in CI meteorites of 5.50 $\pm$ 0.03 dex
(Lodders \cite{Lodders03}).

The atomic parameters influence neither the line profiles nor the derived
abundances significantly. The assumption of hydrogenic photoionization
cross-sections leads to an overestimation of the abundance by 0.04 dex in
comparison to the case when the enhancement factor of 1000 is applied. Despite
the importance of the HFS, no significant errors associated with the
\emph{choice} of HFS data are expected in the analysis. The 0.5 mK accuracy
of measurements produces an uncertainty of 0.01 dex in the abundance. Anstee
$\&$ O'Mara's damping constants for stronger lines in some multiplets lead to
abundances that are smaller than those obtained from the weaker lines.
Adjustment of $C_{\rm 6}$ values inside each multiplet increases the average
abundance by 0.03 dex. However, the discrepancy of 0.2 dex between the
abundances from lines of multiplets 16, 21 and 22 and all other multiplets still
remains. We can not with confidence attribute this to errors in damping or in
the oscillator strengths. Two findings provide evidence for the latter: (1)
correlation of abundances with the Oxford oscillator strengths and (2) similar
regularity found by Simmons $\&$ Blackwell et al. (\cite{Simmons82}) and
Blackwell et al. (\cite{Black82}) in the analysis of solar \ion{Fe}{i} lines
with Oxford $\log gf$ values. Collisions with neutral hydrogen atoms are able to
thermalize the levels only if an enhancement factor $\SH > 10$, with respect to
Drawin's cross-sections, is applied. For the smaller enhancement factors,
although the differences between NLTE and LTE profiles decrease, the abundance
corrections do not exceed 0.05 dex. Total neglect of collisions with hydrogen
requires a reduction of the abundance by 0.01 dex.

The main source of errors in this work is the uncertainties of measured
oscillator strengths, which are inaccurate within a range of 0.04 to 0.1 dex.
However, only the product of abundance and oscillator strengths $\loggfe$ will
be used for studies of manganese in metal-poor stars, where deviations from LTE
are expected to be large.
\begin {acknowledgements}
MB acknowledges with gratitude the Max-Planck Institute for Extraterrestrial
Physics (Germany) and IMPRS-Marie Curie Training Site for her PhD fellowship. MB
thanks L. I. Mashonkina for valuable comments and discussions.
\end {acknowledgements}
%

\Online
\begin{figure*}
\centering
\resizebox{\textwidth}{!}{\rotatebox{90}{\includegraphics{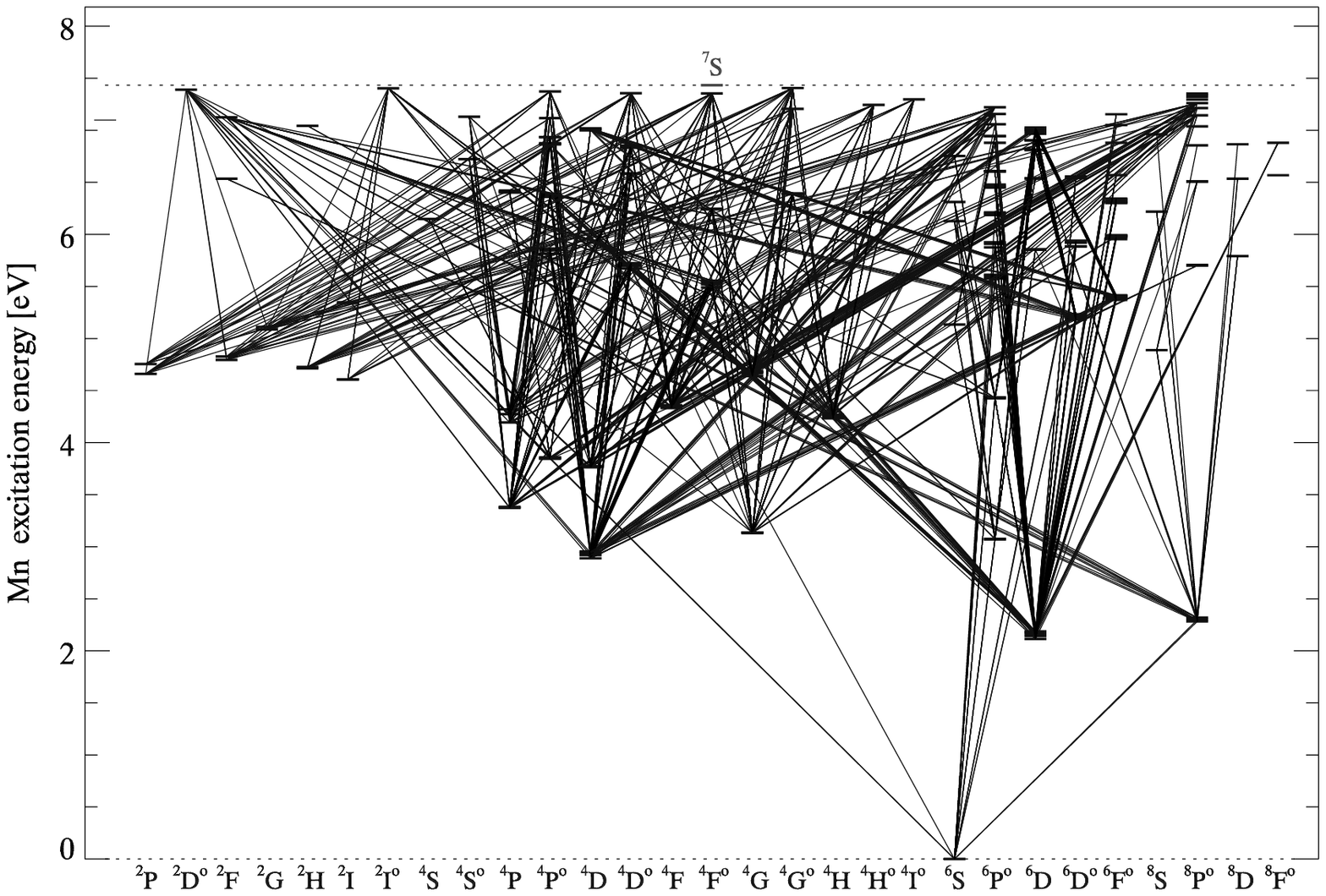}}}
\caption[]{Grotrian diagram of the $\ion{Mn}{i}$ model atom. Solid lines
represent allowed and forbidden transitions included in the model atom.}
\label{atom}
\end{figure*}
\begin{table*}
\begin{minipage}{\linewidth}
\centering
\renewcommand{\footnoterule}{}  
\caption{HFS constants $A$ and $B$ (in units of $10^{-3}$ cm$^{-1}$)
representing magnetic dipole and electric quadrupole interactions for
\ion{Mn}{i} levels. Level energies $E$ are given in eV.}\label{HFS}
\tabcolsep1.5mm
\begin{center}
\begin{tabular}{rlrrrrc|rlrrrrc}
\noalign{\smallskip}\hline\noalign{\smallskip} No. &  level &  $g$  &  $E$ & $A$
&   $B$   &  source & No. &  level &  $g$  &  $E$ &   $A$   &   $B$   &
 source\footnote{References: ~~~(1) Davis et al. \cite{Davis71}; (2)
Dembczy\'nsky et al. \cite{Demb79}; (3) Brodzinski et al. \cite{Brodz87}; (4)
Blackwell-Whitehead et al. \cite{Bl-Wh05};\\ (5) Handrich et al.
\cite{Handrich69}; (6) Johann et al. \cite{Johann81}; (7) Luc \& Gerstenkorn
\cite{Luc72}; (8) Lef\`ebvre et al. \cite{Lefebvre03}; (9) Ba\c{s}ar et al.
\cite{Basar03}}\\
\noalign{\smallskip}\hline\noalign{\smallskip}
  1  &   \Mn{a}{6}{S}{ }{5/2} &  5  &  0.000 &   -2.4 &   0.0   & 1  &  34  &
\Mn{z}{6}{F}{o}{11/2} & 11 &  5.370 &   6.~~ &   0.0   &  8 \\
  2  &   \Mn{a}{6}{D}{ }{9/2} &  9  &  2.110 &   17.~~~&   4.4   & 2  &  35  &
\Mn{z}{6}{F}{o}{9/2} &  9  &  5.384 &   5.2  &   0.0   &  8 \\
  3  &   \Mn{a}{6}{D}{ }{7/2} &  7  &  2.140 &   15.3 &   0.7   & 2  &  36  &
\Mn{z}{6}{F}{o}{7/2} &  7  &  5.396 &   4.6  &   0.0   &  8 \\
  4  &   \Mn{a}{6}{D}{ }{5/2} &  5  &  2.160 &   14.6 & --1.5   & 2  &  37  &
\Mn{z}{6}{F}{o}{5/2} &  5  &  5.405 &   5.8  &   0.0   &  8 \\
  5  &   \Mn{a}{6}{D}{ }{3/2} &  3  &  2.180 &   15.7 & --2.2   & 2  &  38  &
\Mn{z}{6}{F}{o}{3/2} &  3  &  5.411 &   9.1  &   0.0   &  8 \\
  6  &   \Mn{a}{6}{D}{ }{1/2} &  1  &  2.190 &   29.4 &   0.0   & 2  &  39  &
\Mn{z}{4}{F}{o}{9/2} &  9  &  5.490 &   4.4  &   0.0   &  4 \\
  7  &   \Mn{z}{8}{P}{o}{5/2} &  5  &  2.280 &   19.1 &   0.9   & 3  &  40  &
\Mn{z}{4}{F}{o}{7/2} &  7  &  5.520 &   5.7  &   0.0   &  4 \\
  8  &   \Mn{z}{8}{P}{o}{7/2} &  7  &  2.300 &   18.2 & --3.4   & 3  &  41  &
\Mn{z}{4}{F}{o}{5/2} &  5  &  5.540 &   9.5  &   0.0   &  4 \\
  9  &   \Mn{z}{8}{P}{o}{9/2} &  9  &  2.320 &   15.2 &   1.6   & 3  &  42  &
\Mn{z}{4}{F}{o}{3/2} &  3  &  5.560 &  22.3  &   0.0   &  4 \\
 10  &   \Mn{a}{4}{D}{ }{7/2} &  7  &  2.890 &  --5.4 &   0.0   & 4  &  43  &
\Mn{x}{6}{P}{o}{7/2} &  7  &  5.578 &   9.4  &   0.0   &  4 \\
 11  &   \Mn{a}{4}{D}{ }{5/2} &  5  &  2.920 &  --4.6 &   0.0   & 4  &  44  &
\Mn{x}{6}{P}{o}{5/2} &  5  &  5.599 &   9.7  &   0.0   &  4 \\
 12  &   \Mn{a}{4}{D}{ }{3/2} &  3  &  2.940 &    1.7 &   0.0   & 4  &  45  &
\Mn{x}{6}{P}{o}{3/2} &  3  &  5.611 &  12.5  &   0.0   &  4 \\
 13  &   \Mn{a}{4}{D}{ }{1/2} &  1  &  2.950 &   50.6 &   0.0   & 4  &  46  &
\Mn{z}{4}{D}{o}{7/2} &  7  &  5.670 &   1.5  &   0.0   &  9 \\
 14  &   \Mn{z}{6}{P}{o}{3/2} &  3  &  3.070 &   19.1 &   0.4   & 5  &  47  &
\Mn{z}{4}{D}{o}{5/2} &  5  &  5.700 &   2.7  &   0.0   &  9 \\
 15  &   \Mn{z}{6}{P}{o}{5/2} &  5  &  3.070 &   15.6 & --2.5   & 5  &  48  &
\Mn{z}{4}{D}{o}{3/2} &  3  &  5.710 &   6.4  &   0.0   &  9 \\
 16  &   \Mn{z}{6}{P}{o}{7/2} &  7  &  3.080 &   14.3 &   2.1   & 5  &  49  &
\Mn{z}{4}{D}{o}{1/2} &  1  &  5.730 &  35.~~ &   0.0   &  9 \\
 17  &   \Mn{a}{4}{G}{ }{11/2}& 11  &  3.133 &   13.5 &   0.0   & 6  &  50  &
\Mn{e}{8}{D}{ }{3/2} &  3  &  5.791 &  38.4  &   0.0   &  4 \\
 18  &   \Mn{a}{4}{G}{ }{9/2} &  9  &  3.135 &   13.2 &   0.0   & 6  &  51  &
\Mn{e}{8}{D}{ }{5/2} &  5  &  5.791 &  24.~~ &   0.0   &  4 \\
 19  &   \Mn{a}{4}{G}{ }{7/2} &  7  &  3.135 &   14.6 &   0.0   & 6  &  52  &
\Mn{e}{8}{D}{ }{7/2} &  7  &  5.791 &  17.6  &   0.0   &  4 \\
 20  &   \Mn{a}{4}{G}{ }{5/2} &  5  &  3.134 &   19.9 &   0.0   & 6  &  53  &
\Mn{e}{8}{D}{ }{9/2} &  9  &  5.791 &  15.7  &   0.0   &  4 \\
 21  &   \Mn{z}{4}{P}{o}{5/2} &  5  &  3.840 & --20.3 &   2.5   & 3  &  54  &
\Mn{e}{8}{D}{ }{11/2} & 11 &  5.792 &  14.5  &   0.0   &  4 \\
 22  &   \Mn{z}{4}{P}{o}{3/2} &  3  &  3.850 & --27.1 & --1.3   & 3  &  55  &
\Mn{y}{4}{P}{o}{5/2} &  5  &  5.820 & --1.~~ &   0.0   &  4 \\
 23  &   \Mn{z}{4}{P}{o}{1/2} &  1  &  3.860 & --71.1 &   0.0   & 3  &  56  &
\Mn{y}{4}{P}{o}{3/2} &  3  &  5.850 &--10.~~ &   0.0   &  4 \\
 24  &   \Mn{y}{6}{P}{o}{3/2} &  3  &  4.430 & --32.4 &   0.6   & 7  &  57  &
\Mn{y}{4}{P}{o}{1/2} &  1  &  5.870 &--32.5  &   0.0   &  4 \\
 25  &   \Mn{y}{6}{P}{o}{5/2} &  5  &  4.430 & --18.~~~& --2.3   & 7  &  58  &
\Mn{e}{6}{D}{ }{9/2} &  9  &  5.850 &  15.5  &   0.0   &  7 \\
 26  &   \Mn{y}{6}{P}{o}{7/2} &  7  &  4.440 & --13.~~~& --6.2   & 7  &  59  &
\Mn{e}{6}{D}{ }{7/2} &  7  &  5.850 &  15.8  &   0.0   &  7 \\
 27  &   \Mn{e}{8}{S}{ }{7/2} &  7  &  4.890 &   24.6 &   1.6   & 3  &  60  &
\Mn{e}{6}{D}{ }{5/2} &  5  &  5.850 &  17.6  &   0.0   &  7 \\
 28  &   \Mn{e}{6}{S}{ }{5/2} &  5  &  5.130 &   27.~~~&   0.0   & 3  &  61  &
\Mn{e}{6}{D}{ }{3/2} &  3  &  5.860 &  22.8 &   0.0   &  7 \\
 29  &   \Mn{z}{6}{D}{o}{9/2} &  9  &  5.180 &    2.9 &   0.0   & 4  &  62  &
\Mn{e}{6}{D}{ }{1/2} &  1  &  5.860 &  61.6  &   0.0   &  7 \\
 30  &   \Mn{z}{6}{D}{o}{7/2} &  7  &  5.200 &    1.3 &   0.0   & 4  &  63  &
\Mn{f}{6}{S}{ }{5/2} &  5  &  7.024 &--20.6  &   0.0   &  8 \\
 31  &   \Mn{z}{6}{D}{o}{5/2} &  5  &  5.210 &  --0.9 &   0.0   & 4  &  64  &
\Mn{e}{4}{S}{ }{3/2} &  3  &  7.119 &--50.5  &   0.0   &  8 \\
 32  &   \Mn{z}{6}{D}{o}{3/2} &  3  &  5.230 &  --5.~~~&   0.0   & 4  &  65  &
\Mn{g}{6}{S}{ }{5/2} &  5  &  6.311 &  23.2  &   0.0   &  8 \\
 33  &   \Mn{z}{6}{D}{o}{1/2} &  1  &  5.230 & --27.4 &   0.0   & 4  &      &
 &     &    &    &    & \\
\noalign{\smallskip}\hline\noalign{\smallskip}
\end{tabular}
\end{center}
\end{minipage}
\end{table*}
%
%
\end{document}